\documentclass[twocolumn,aps,showpacs,pre,amsmath,amssymb,floatfix,longbibliography]{revtex4-2}

\usepackage{color}
\usepackage{multirow}
\usepackage{bm}
\usepackage{epsfig}
\usepackage{psfrag}
\usepackage[ansinew]{inputenc}
\usepackage[english]{babel} 
\usepackage{amsfonts}
\usepackage{amsmath}
\usepackage{upgreek}
\usepackage{enumitem}
\usepackage[dvipsnames]{xcolor}
\usepackage{environ}  
\usepackage{hyperref}
\usepackage[extra]{tipa}
\usepackage[bb=libus]{mathalpha}

\begin{document}

\title{
	Unified Description of Learning Dynamics in the Soft Committee Machine from Finite to Ultra-Wide Regimes
	}

\author{
Assem Afanah}
\affiliation{
	Institut f\"{u}r Theoretische Physik, Universit\"{a}t
  Leipzig,  Br\"{u}derstrasse 16, 04103 Leipzig, Germany
	}

\author{
Bernd Rosenow}
\affiliation{
	Institut f\"{u}r Theoretische Physik, Universit\"{a}t
  Leipzig,  Br\"{u}derstrasse 16, 04103 Leipzig, Germany
	}	
\date{\today}

\begin{abstract}

We study the learning dynamics of the soft committee machine (SCM) with Rectified Linear Unit (ReLU) activation using a statistical-mechanics approach within the annealed approximation. The SCM consists of a student network with $N$ input units and $K$ hidden units trained to reproduce the output of a teacher network with $M$ hidden units. We introduce a reduced set of macroscopic order parameters that yields a unified description valid from the conventional regime $K \ll N$ to the ultra-wide limit $K \ge N$. 
The control parameter $\alpha$, proportional to the ratio of training samples to adjustable weights, serves as an effective measure of dataset size. 
 For small $\gamma = M/N$, we recover a continuous phase transition at $\alpha_{c} \approx 2\pi$ from an unspecialized, permutation-symmetric state to a specialized state in which student units align with the teacher. For finite $\gamma$, the transition disappears and the generalization error decreases smoothly with dataset size, reaching a low plateau when $\gamma=1$. In the asymptotic limit $\alpha \to \infty$, the error scales as $\varepsilon_{g} \propto 1/\alpha$, independent of $\gamma$ and $K$. The results highlight the central role of network dimensions in SCM learning and provide a framework extendable to other activations and quenched analyses.

\end{abstract}
 
\maketitle
\section{Introduction}
\begin{figure}[t]
	\centering
	\includegraphics[width=8.6cm]{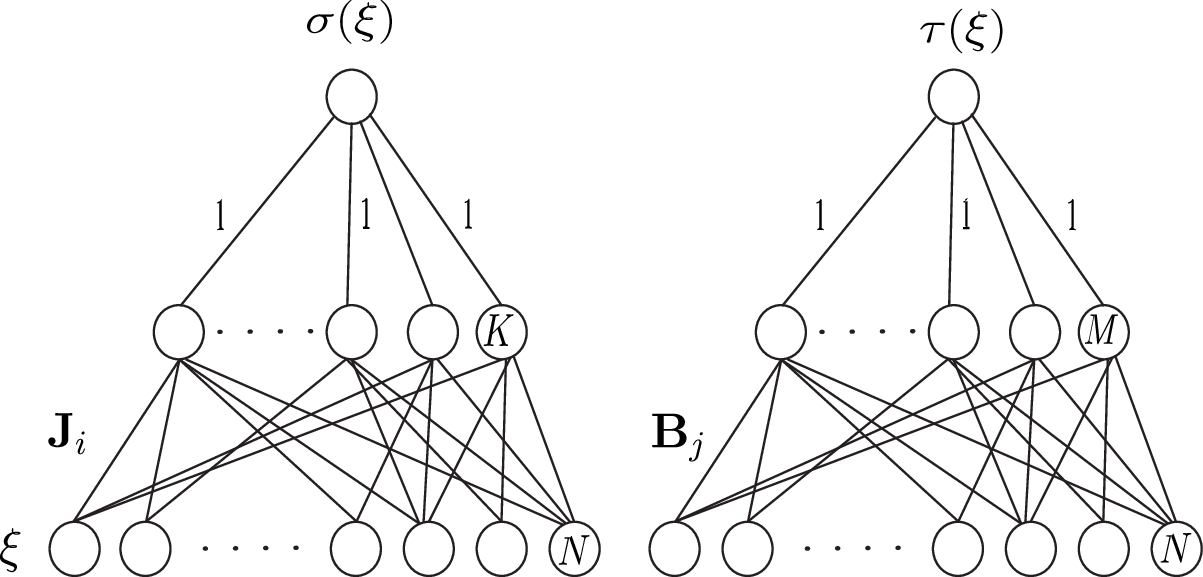}
	\caption{\label{Fig:SCM}
Schematic diagrams of the student and teacher soft committee machines. 
Both networks receive an $N$-dimensional input and contain $M$ (teacher) or $K$ (student) hidden units. The corresponding input-hidden weight vectors are denoted by ${\bm{B}}_{j}$ for the teacher and ${\bm{J}}_{i}$ for the student; the input-hidden weight vectors are normalized to one.
For a given input ${\bm{\xi}} \in \mathbb{R}^{N}$, the outputs of the teacher, $\tau({\bm{\xi}})$, and of the student, $\sigma({\bm{\xi}})$, are proportional to the sum of hidden-unit activations under a Rectified Linear Unit (ReLU) activation, $g(x) = x \Theta(x)$, where $\Theta(x)$ is the Heaviside step function.  }
\end{figure}

Neural networks have long been a subject of intensive experimental and theoretical investigation \cite{engel,book_shallow,phys_Mash_rev,Neal1996,goodfellow2016}. 
Despite remarkable technological progress \cite{COLLINS2021,niskanen2023,Deepbook}, a complete theoretical  understanding of their behavior remains a challenge. 
Physicists have applied methods from statistical mechanics -- such as dynamical mean-field theory and the replica method, originally developed for spin glasses and disordered systems -- to characterize the complex dynamics of neural networks \cite{spinNN,phase_Bahri,Sompolinsky1988,Gabrie2020,MEHTA20191,RevModPhys.65.499}.  Early work on perceptron learning introduced a framework in which a small set of order parameters describes the generalization behavior of neural networks in the thermodynamic limit \cite{gardner,Tishby,Gardner1988}.
These approaches were later extended to multilayer networks with diverse architectures and activation functions \cite{Levin,PhysRevLett.87.078101,RUrbanczik_1995,saad}.

Here we focus on the generalization behavior of the soft committee machine (SCM) \cite{saad,Saad95prl}, a two-layer network with a single output unit whose response is the average of its hidden-unit activations (see Fig.~\ref{Fig:SCM}). The SCM is typically studied in a student-teacher setting, where a student network with $N$ inputs and $K$ hidden units attempts to reproduce the output of a teacher network with $M$ hidden units.
In this framework, ${\bm{J}}_{i}$ denotes the normalized adaptive weight vectors of the student, while ${\bm{B}}_{j}$ represents orthonormal weight vectors of the teacher, and we employ the ReLU activation function \cite{nair2010}. Recent interest has shifted to ultra-wide networks with $K \geq N$, and even to the infinite-width limit, motivated by empirical observations that such systems often display improved generalization \cite{PhysRevLett.87.078101,doubledecent,Rosen}.
In this limit, the network becomes formally equivalent to a Gaussian process via the neural tangent kernel (NTK) \cite{lee2018,Jacot2018,Arora2019,AllenZhu2019}, offering a tractable route toward understanding training dynamics in high-dimensional regimes \cite{Advani2020,Li2021}.

A conventional statistical-mechanics treatment of the SCM involves $\mathcal{O}(K^2)$ order parameters. 
Typical properties follow from evaluating the free energy as a function of these order parameters, using either the annealed approximation \cite{H.Schwarze_1993,HSchwarze2_1993} or the more accurate (but technically demanding) quenched-average approach \cite{HSchwarze_1993,Ahr_1999}, based on the replica method \cite{Mezard1987}. In the high-temperature limit, the annealed approximation becomes exact and coincides with the quenched description, providing a convenient framework to explore the SCM under various learning scenarios. However, this conventional formalism breaks down in the ultra-wide regime: the number of order parameters then exceeds the actual number of degrees of freedom, in conflict with the notion that an order parameter should represent a macroscopic property of an ensemble of microstates.

 To address this issue, we develop a formulation that depends explicitly on $(N,K,M)$, allowing us to find a unified description  of the SCM that remains valid even when $K \geq N$ assuming $M \ll N$. 
Following the standard approach, one  defines the 
  self-averaging quantities $Q_{ij} =\bm{J}_{i} \cdot\bm{J}_{j} / N $ and $R_{ij} =\bm{J}_{i} \cdot\bm{B}_{j} / N $   as  macroscopic order parameters. Here, instead, we introduce a reduced set of parameters  that average over the contributions of individual student and teacher units overlaps:
\begin{align}
\nonumber \tilde{Q} = \dfrac{M}{K^{2}}  \sum_{i,j=1}^{K} & Q_{ij} \quad,  \quad \tilde{R} = \dfrac{1}{K} \sum_{i=1}^{K} \sum_{j=1}^{M} R_{ij} \\
& \tilde{r} = \dfrac{1}{K} \sum_{i=1}^{K} \sum_{j=1}^{M} R^{2}_{ij} ~ . \label{Eq:order} 
\end{align}   
The annealed free energy Eq.~(\ref{Eq:free1}) can be expressed in terms of $(\tilde{Q}, \tilde{R}, \tilde{r})$. 
Minimizing the free energy with respect to these parameters yields the generalization error $\varepsilon_{g}$ at the saddle point as a function of the number of training examples $\alpha$ (rescaled by the input dimension and the number of hidden units).  Details of the derivation and the explicit form of $\varepsilon_{g}$ are presented in the Model section.

\begin{figure}[t]
    \centering
	\includegraphics [width=\linewidth]{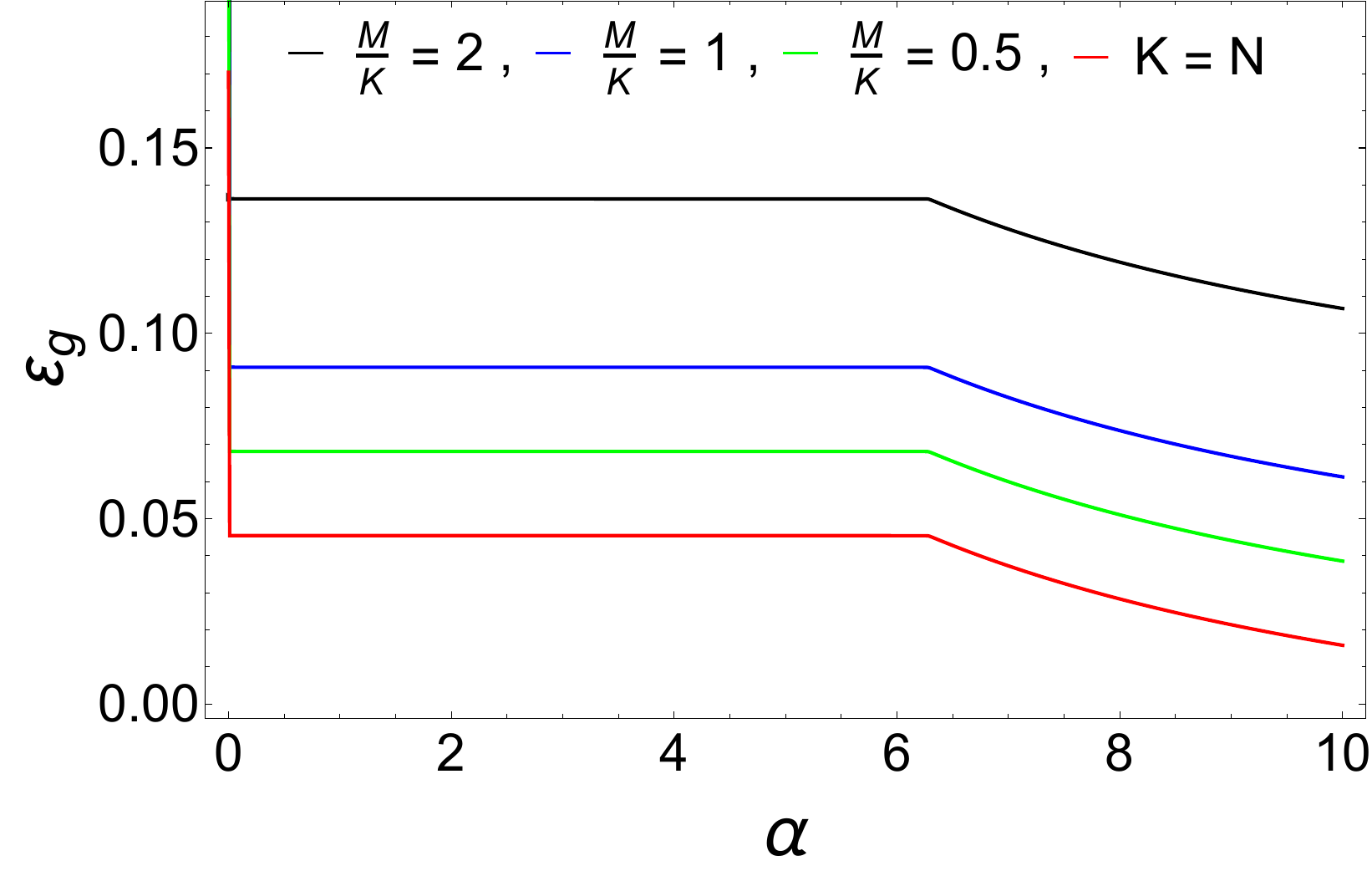}
	\caption{ Learning curves obtained by numerically minimizing the free energy, Eq.~(\ref{Eq:free1}),  for $(N = 10^{12}, M = 10^{6})$ and various ratios  $M/K$. The unrealizable ($M/K=2$), realizable ($M/K=1$), over-realizable ($M/K=0.5$), and ultra-wide ($K \geq N$) regimes all display a phase transition from an unspecialized to a specialized phase near $\alpha_{c} \approx 2\pi$. The qualitative form of the learning curves remains similar across these regimes; only the height of the symmetric plateau decreases with increasing $K/M$. Once $K = N$, the plateau height saturates and no further decrease is observed. }
\label{Fig:KLM} 
\end{figure}

\begin{figure}[t]
	\centering
	\includegraphics [width=\linewidth]{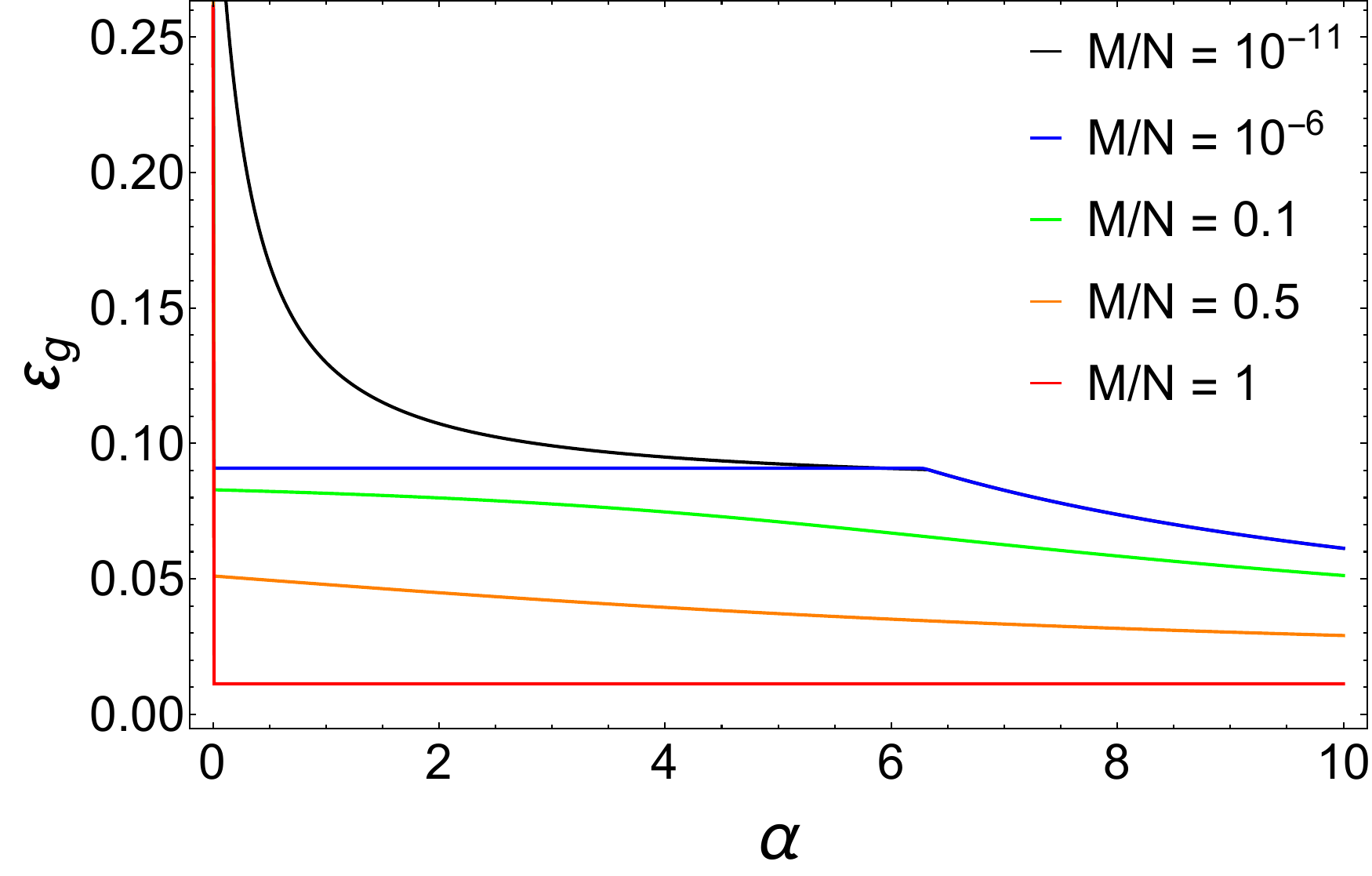}
	\caption{ Evolution of the learning curves as $M/N$ varies in the realizable case $K=M$ with $N=10^{12}$. For $M/N \ll 1$ (e.g., $M/N = 10^{-11}$ or $10^{-6}$), a phase transition occurs at $\alpha_{c} \approx 2\pi$. For large networks with $K(M) \gg 1$ hidden units (blue curve, $K = 10^{6}$), a well-defined symmetric plateau develops.
When $10^{-3} \lesssim M/N \lesssim 1$, the generalization error decreases smoothly with $\alpha$, and no sharp transition is observed (shown for $M/N = 0.1$ and $0.5$). At $M/N = 1$, the generalization error immediately reaches a low, $\alpha$-independent plateau. }
\label{Fig:KeqM_flatt} 
\end{figure}

Representative results are shown in Figs.~\ref{Fig:KLM} and~\ref{Fig:KeqM_flatt},  obtained by numierically minimizing the free energy for different learning scenarios with $M \ll N$. In Fig.~\ref{Fig:KLM}, we  examine the realizable  ($K=M$), unrealizable ($K<M$), over-realizable  ($K>M$), and  ultra-wide  ($K \geq N$) cases.
The learning curves exhibit a qualitatively similar structure across these regimes:  the generalization error $\varepsilon_{g}$ decreases rapidly to a plateau corresponding to an unspecialized state in which the hidden units of the student are permutation symmetric. This symmetry is broken at a critical value $\alpha_{c}$, where a second-order phase transition leads to a specialized state in which the student gradually aligns with the teacher. 
The plateau height depends on $K/N$, reaching its minimum when $K \geq N$ (red curve in Fig.~\ref{Fig:KLM}).

The existence of this phase transition has been reported in various SCM models with different learning rules and activation functions \cite{Oostwal,Biehl_1998,saad,Richert_2022}. Figure~\ref{Fig:KeqM_flatt} shows that, for the realizable case $K = M$, the transition is not universal: whether a distinct symmetric plateau appears, and the detailed nature of the learning trajectory, depend sensitively on $M/N$.

For $M/N \ll 1$, the network undergoes a transition near $\alpha_{c} \approx 2\pi$. A well-defined symmetric plateau arises only for networks with $K(M) \rightarrow \infty$, where corrections of $\mathcal{O}(1/K)$ can be neglected. For finite $K$, these corrections contribute significantly, producing a smooth decrease of $\varepsilon_{g}$ in the symmetric phase while retaining a kink at $\alpha = \alpha_c$.
When $M/N$ is finite, no sharp transition is observed: the generalization error decreases continuously with $\alpha$ after an initial drop, and for $M/N = 1$, it settles immediately at the lowest plateau value, independent of $\alpha$.

 These  findings highlight the importance of properly accounting for the network dimensions $(N,K,M)$ in analyzing SCM-based models. 
 In particular, when $K$ becomes large relative to $N$, the standard statistical-mechanics formalism fails to capture the true behavior of the system. The implications of this breakdown, and possible extensions of the framework, are discussed in the concluding section.


\section{Model} 

We study the SCM in a student-teacher setup, where a student network with $K$ hidden units is trained to reproduce the rule implemented by a teacher network with $M$ hidden units. The activation function is the Rectified Linear Unit (ReLU), introduced by Nair and Hinton \cite{nair2010} and widely adopted in modern deep learning for its rapid convergence and superior generalization relative to sigmoidal functions \cite{relu_speech,relu_empirical}.
For an input vector ${\bm{\xi}} \in \mathbb{R}^{N}$, the outputs of the student and teacher networks are
\small 
\begin{align}
\sigma = \dfrac{\sqrt{M}}{K} \sum_{i=1}^{K} g\left( \dfrac{1}{\sqrt{N}}\bm{J}_{i} \cdot \bm{\xi}^{\mu} \right) ,~ \tau = \dfrac{1}{\sqrt{M}} \sum_{j=1}^{M} g\left( \dfrac{1}{\sqrt{N}}\bm{B}_{j} \cdot \bm{\xi}^{\mu}\right) \ ,
\end{align}
\normalsize
where $g(x) = x\,\Theta(x)$ is the ReLU function. 
The student's adaptive weight vectors $\{{\bm{J}}_{i}\}$ satisfy ${\bm{J}}_{i}^{2} = N$, while the teacher's weight vectors ${{\bm{B}}{j}}$ are orthonormal,
${\bm{B}}{i}\cdot\bm{B}{j} = N \,\delta_{ij}$.

The student is trained on a dataset
$\mathbb{D}= \left\lbrace \bm{\xi}^{\mu}, \tau(\bm{\xi}^{\mu}) \right\rbrace$ with $\mu=1,2,..,P$ of random i.i.d.~inputs with unit variance per component. Its performance is measured by the quadratic cost function
\begin{align}
\epsilon_{t} = \dfrac{1}{P} \sum_{\mu = 1}^{P} \frac{1}{2} \left[ \sigma({\bm{\xi}}^{\mu}) - \tau({\bm{\xi}}^{\mu})\right]^{2}\ .
\end{align}
The generalization error, which quantifies the expected performance on unseen inputs, is
\begin{align}
\varepsilon_{g} = \dfrac{1}{2} \left\langle \left[ \dfrac{\sqrt{M}}{K} \sum_{i=1}^{K} g(x_{i}) - \dfrac{1}{\sqrt{M}} \sum_{j=1}^{M} g(y_{j})\right]^{2}  \right\rangle_{\bm{\xi}} \ ,  \label{Eq:EgAvg} 
\end{align}
where the average $\langle \cdot \rangle_{\bm{\xi}}$ is taken over the distribution of random inputs. 
We define the local fields  $x_{i} = {\bm{J}_{i}} \cdot {\bm{\xi}} / \sqrt{N}$ and $y_{j} = {\bm{B}_{j}} \cdot\bm{\xi} / \sqrt{N}$. 
If the components of $\underline{\bm{\xi}}$ are drawn i.i.d.~from a Gaussian distribution with zero mean and unit variance, then in the limit $N \to \infty$ the central limit theorem implies that the joint distribution $\mathcal{P}(\bm{x},\bm{y})$ of ${x_i}$ and ${y_j}$ is Gaussian, with moments \cite{Biehl_1998,PhysRevE.47.1384} 
\begin{align}
\nonumber \left\langle x_{i} \right\rangle = 0, ~ \left\langle x_{i} ~ x_{j} \right\rangle &= Q_{ij},~ \left\langle y_{i} \right\rangle = 0,~ \left\langle y_{i} y_{j} \right\rangle = T_{ij}, \\
  & \left\langle x_{i} y_{j} \right\rangle = R_{ij} \label{Eq:moments} 
\end{align} 
where $Q_{ij}$ and $R_{ij}$ denote the student-student and student-teacher overlaps, respectively, and
$T_{ij} = {\bm{B}}_{i}\cdot {\bm{B}}_{j}/N$ is the teacher-teacher overlap.
From these moments, the generalization error can be computed exactly \cite{Oostwal} as
\begin{align}
\nonumber \varepsilon_{g} &=  \, \dfrac{M}{2 K^{2}} \, \sum_{i,j=1}^{K} \left( \dfrac{Q_{ij}}{4} + \dfrac{\sqrt{1- Q^{2}_{ij}}}{2 \pi} + \dfrac{Q_{ij} \arcsin [Q_{ij}]}{2 \pi} \right) \\
\nonumber & - \dfrac{1}{K} \sum_{i=1}^{K} \sum_{j=1}^{M} \left( \dfrac{R_{ij}}{4} + \dfrac{\sqrt{1- R^{2}_{ij}}}{2 \pi} + \dfrac{R_{ij} \arcsin [R_{ij}]}{2 \pi} \right) \\
& + \dfrac{1}{2 M} \, \sum_{i,j=1}^{M} \left( \dfrac{T_{ij}}{4} + \dfrac{\sqrt{1- T^{2}_{ij}}}{2 \pi} + \dfrac{T_{ij} \arcsin [T_{ij}]}{2 \pi} \right)\  .   \label{Eq:Eg_ost}  
\end{align}

Following the statistical-mechanics formalism, we consider a Gibbs ensemble of student networks with density
$\exp(-\beta P \epsilon_{t})/Z$,
where the training error acts as an energy term, $P$ denotes the number of training examples,  and the inverse temperature $\beta = 1/T$ controls the thermal noise.
The partition function is
\begin{align}
Z = \int \prod_{i=1}^{K} ~ d\mu({\bm{J}}_{i}) \exp(-\beta P \epsilon_{t})
\end{align} 
which integrates over all normalized student weight configurations.
The measure $d\mu({\bm{J}}_{i})$ enforces normalization of each weight vector.
Typical system properties follow from the quenched free energy
\begin{align}
- \beta F = \left\langle  \text{ln} Z \right\rangle \ .
\end{align}
Evaluating this average is generally intractable and requires the application of replica method. However, in the high-temperature limit $\beta \to 0$, the annealed approximation
$\langle \ln Z \rangle \approx \ln \langle Z \rangle$ becomes exact and simplifies the free energy calculation with
\begin{align}
\nonumber \left\langle  Z \right\rangle  =& \int \prod_{i=1}^{K} ~ d\mu({\bm{J}}_{i}) \exp(-\beta P <\epsilon_{t}>) \\
                =& \int \prod_{i=1}^{K} ~ d\mu({\bm{J}}_{i}) \exp(-\beta P \varepsilon_{g}).
\end{align}
In this formulation, $Q_{ij}$ and $R_{ij}$ act as macroscopic order parameters, while the orthonormal teacher vectors give $T_{ij} = \delta_{ij}$, contributing only a constant to the free energy.
For $N \gg K$, there are $K(K-1)/2$ independent $Q_{ij}$ and $MK$ $R_{ij}$ parameters, with $Q_{ii} = 1$.
When $K \geq N$, however, the number of order parameters exceeds the number of degrees of freedom, rendering the standard description inconsistent.
We therefore introduce an alternative formulation that remains valid in both regimes.
Expanding the nonlinear terms in Eq.~(\ref{Eq:Eg_ost}) to second order in ${Q_{ij}, R_{ij}}$ gives
\begin{align}
\nonumber \varepsilon_{g} \approx \, & \dfrac{M}{2 K^{2}} \, \left( \sum_{i,j=1}^{K} \dfrac{Q_{ij}}{4}
 + \sum_{i\neq j}^{K}\dfrac{Q^{2}_{ij}}{4 \pi} \right)\\
\nonumber  &- \dfrac{1}{K} \sum_{i=1}^{K} \sum_{j=1}^{M} \left( \dfrac{R_{ij}}{4} +\dfrac{R^{2}_{ij}}{4 \pi} \right) \\
           &  + \dfrac{M}{K} \left(\frac{1}{8}- \frac{1}{4 \pi} \right) + \left( \dfrac{1}{4}- \dfrac{1}{4 \pi} \right). \label{Eq:Eg_expnd}  
\end{align}
For large $N$, random vectors are nearly orthogonal \cite{Andrecut_2018}, implying
$\sum_{ij}Q_{ij} \gg \sum_{i\neq j}Q_{ij}^{2}$.
We therefore retain only the dominant linear terms in $Q_{ij}$, while keeping higher-order terms in $R_{ij}$ to capture the transition to the specialized phase.
Introducing the aggregated order parameters of Eq.~(\ref{Eq:order}), we rewrite the generalization error as
\begin{align}
\nonumber \varepsilon_{g}(\tilde{Q},\tilde{R},\tilde{r}) = \, & \dfrac{\tilde{Q}}{8} -  \dfrac{\tilde{R}}{4} - \dfrac{\tilde{r}}{4 \pi} +
 \dfrac{M}{K} \left(\frac{1}{8}- \frac{1}{4 \pi} \right) \\ 
&+ \left( \dfrac{1}{4}- \dfrac{1}{4 \pi} \right) \ , \label{Eq:Eg_QRr} 
\end{align} 
We then introduce these order parameters into the partition function via delta functions, yielding
\begin{align}
 \left\langle  Z \right\rangle = \int ~ d\tilde{Q} ~ d\tilde{R} ~ d\tilde{r} ~ \exp[- N (\alpha K \varepsilon_{g} - S)] \ ,   \label{Eq:Z_int}
\end{align}
where $\alpha = \beta P / (N K)$ denotes the scaled dataset size, and $S$ is the entropic contribution describing the volume of version-space configurations consistent with $(\tilde{Q},\tilde{R},\tilde{r})$.
In the limits $\beta \to 0$ and $P \to \infty$, $\alpha$ remains of order unity.
For large $N$, the integral in Eq.~(\ref{Eq:Z_int}) can be evaluated via a saddle-point approximation, identifying the exponent as the free energy,
\begin{align}
f = \dfrac{\beta F}{N} = \alpha K \varepsilon_{g} - S  ~.
\label{Eq:free1}
\end{align}

The entropic term $S$ is explicitly
\footnotesize
\begin{align}
\nonumber S &=  \dfrac{1}{N} \, \text{ln} \int \prod_{i=1}^{K} ~ d \mu(\underline{\bm{J}}_{i}) ~ \delta \left( \sum_{ij=1}^{K}\bm{J}_{i} \cdot\bm{J}_{j} - \dfrac{N K^{2}}{M} \tilde{Q} \right)\\
 & \times \delta \left(N K \tilde{R} - \sum_{i=1}^{K} \sum_{j=1}^{M}\bm{J}_{i} \cdot\bm{B}_{j}\right) 
  \times \delta \left(N^{2} K \tilde{r} - \sum_{i=1}^{K} \sum_{j=1}^{M} (\underline{\bm{J}}_{i} \cdot\bm{B}_{j})^{2} \right)  ,  \label{Eq:entrop} 
\end{align} 
\normalsize
which measures the volume in version space occupied by student vectors ${\bm{J}}{i}$ consistent with the given order parameters.
 The integral can be evaluated via another  saddle point integration, yielding 
\small
\begin{align}
\nonumber  & S(\tilde{Q},\tilde{R},\tilde{r}) = \min_{\substack{\hat{\lambda}, \hat{Q} \\ \hat{R},\hat{r}}} \left[  const. + K \hat{\lambda} + K \hat{R} \tilde{R} + K \hat{r} \tilde{r} \right. \\
 \nonumber & \left.  - \dfrac{K^{2}}{M} \hat{Q} \tilde{Q} - \dfrac{(1- \gamma)(K-1)}{2} ~ \text{ln} \hat{\lambda} - \dfrac{\gamma(K-1)}{2} \text{ln} (\hat{\lambda} + \hat{r}) \right. \\
   & \left. - \dfrac{(1-\gamma)}{2} \text{ln} (\hat{\lambda} - K \hat{Q}) - \dfrac{\gamma}{2} \text{ln} (\hat{\lambda} + \hat{r} - K \hat{Q}) + \dfrac{\hat{R}^{2}}{4} \dfrac{K M}{\hat{\lambda} + \hat{r} - K \hat{Q}}
 \right]   ,   \label{Eq:entropy1} 
\end{align}
\normalsize
where the dependence on $(N,K,M)$ arises explicitly from the constraint on $\tilde{r}$, with $\gamma = M/N$ for convenience (see Appendix~A).
The auxiliary variables $(\hat{Q}, \hat{R}, \hat{r}, \hat{\lambda})$ enforce the desired overlap structure of the student weight vectors.

\section{Results and Discussion}

In this section we discuss the results obtained for the SCM under various learning scenarios. For a given choice of parameters $(M,K,\gamma)$, a local minimum in the free-energy landscape is obtained by solving
\begin{equation}    
\dfrac{\partial f}{\partial \tilde{Q}} = \dfrac{\partial f}{\partial \tilde{R}} = \dfrac{\partial f}{\partial \tilde{r}} = 0 \ . 
\end{equation}
Numerical solutions are in general required for the saddle-point equations, although in special cases -- most notably $K=M$ with either $\gamma \ll 1$ or $\gamma=1$ -- one can make analytic progress by eliminating the auxiliary variables $(\hat{\lambda},\hat{Q},\hat{R},\hat{r})$ and rewriting the entropic part in terms of $(\tilde{Q}, \tilde{R}, \tilde{r})$.

Figures~\ref{Fig:KLM} illustrate the learning behavior for $(N=10^{12}, \gamma=10^{-6})$. As noted earlier, we compare different ratios $M/K$ for the unrealizable ($K<M$), realizable ($K=M$), over-realizable ($K>M$), and ultra-wide ($K\ge N$) cases. We observe a second-order phase transition at $\alpha_{c}\approx 2\pi$ for $\gamma \ll 1$, largely independent of $M/K$. 
By contrast, Fig.~\ref{Fig:KeqM_flatt} highlights the absence of a phase transition when $\gamma$ is finite, corroborating the strong dependence on $M/N$. {Although our formalism successfully describes the learning behavior in the unspecialized phase and near the transition point in the specialized phase, it becomes inaccurate  deep in the specialized regime  due to the approximation made in Eq.~(\ref{Eq:Eg_expnd}).}

\begin{figure*}[t!]
    \centering
        \includegraphics[width=15 cm, height=15 cm,keepaspectratio]{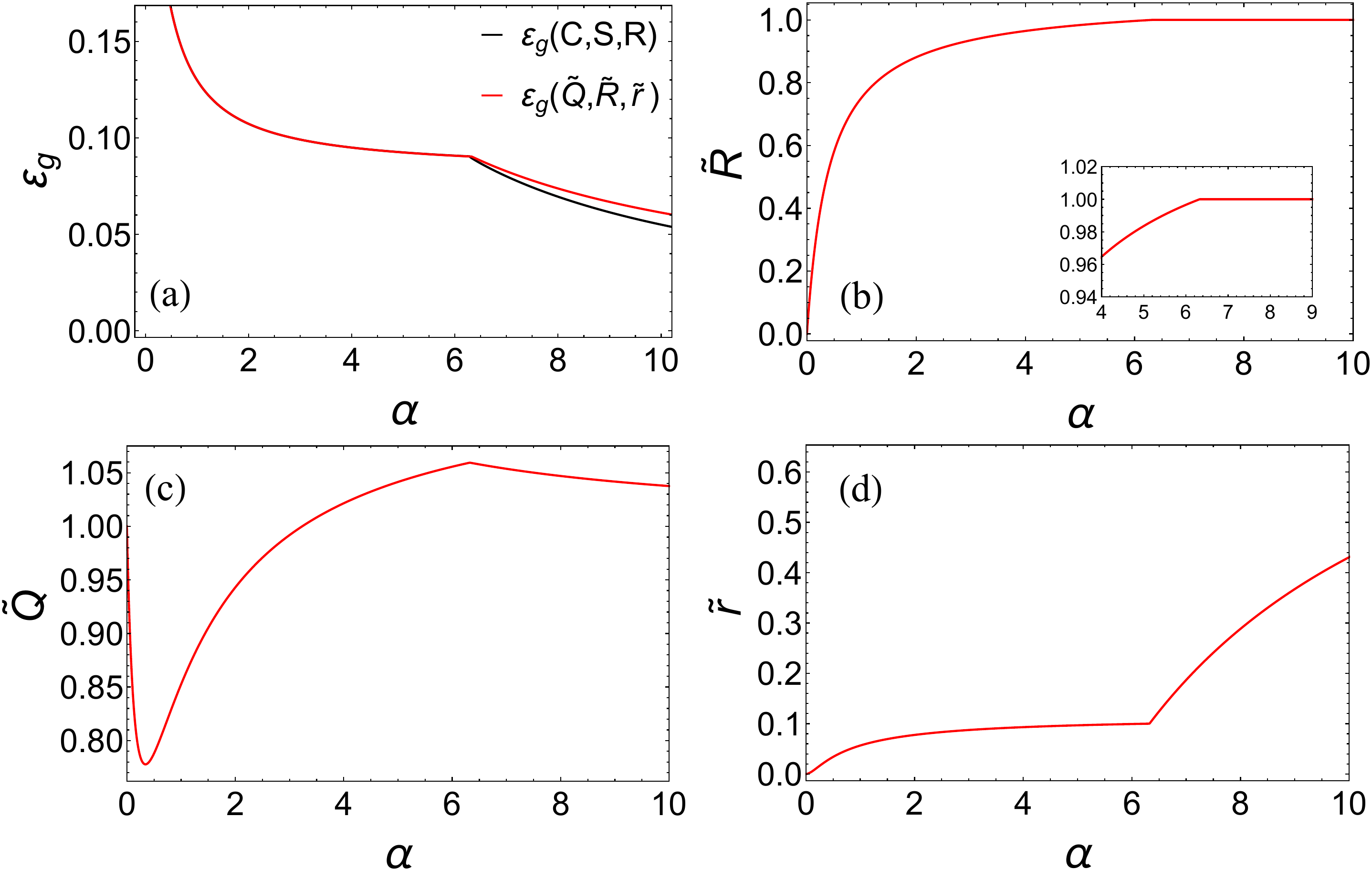}
    \caption{ (a) Generalization error $\varepsilon_{g}(\tilde{Q},\tilde{R},\tilde{r})$ vs.~dataset size $\alpha$ for the realizable case $(K=M)$ with $(\gamma=10^{-11},K=10)$, obtained from minimizing Eq.~(\ref{Eq:free_KeM}).
We compare $\varepsilon_{g}(\tilde{Q},\tilde{R},\tilde{r})$ to $\varepsilon_{g}(C,R,S)$ reproduced from \cite{Oostwal}, of the generalization behavior of a ReLU-based SCM. The two formalisms agree in the unspecialized phase ($\alpha<\alpha_{c}$) and near the phase boundary $\alpha_{c}\approx2\pi$, but differ deeper in the specialized phase ($\alpha>\alpha_{c}$) due to our expansion in Eq.~(\ref{Eq:Eg_expnd}).
(b) Evolution of $\tilde{R}$ with $\alpha$: it grows smoothly in the unspecialized phase and then rapidly approaches $1$ beyond $\alpha_{c}$ (inset).    
(c) $\tilde{Q}$ decreases at small $\alpha$, then rises to a peak at $\alpha_{c}$, signaling the phase transition.    
(d) For $\alpha<\alpha_{c}$, $\tilde{r}\sim\mathcal{O}(1/K)$ (consistent with committee symmetric $R_{ij}$); for $\alpha>\alpha_{c}$, specialization begins and $\tilde{r}\approx 1-2\pi/\alpha$.    }
\label{Fig:Eg_KeM10}
\end{figure*}
\begin{figure*}[t!]
    \centering
        \includegraphics[width=15 cm, height=15 cm,keepaspectratio]{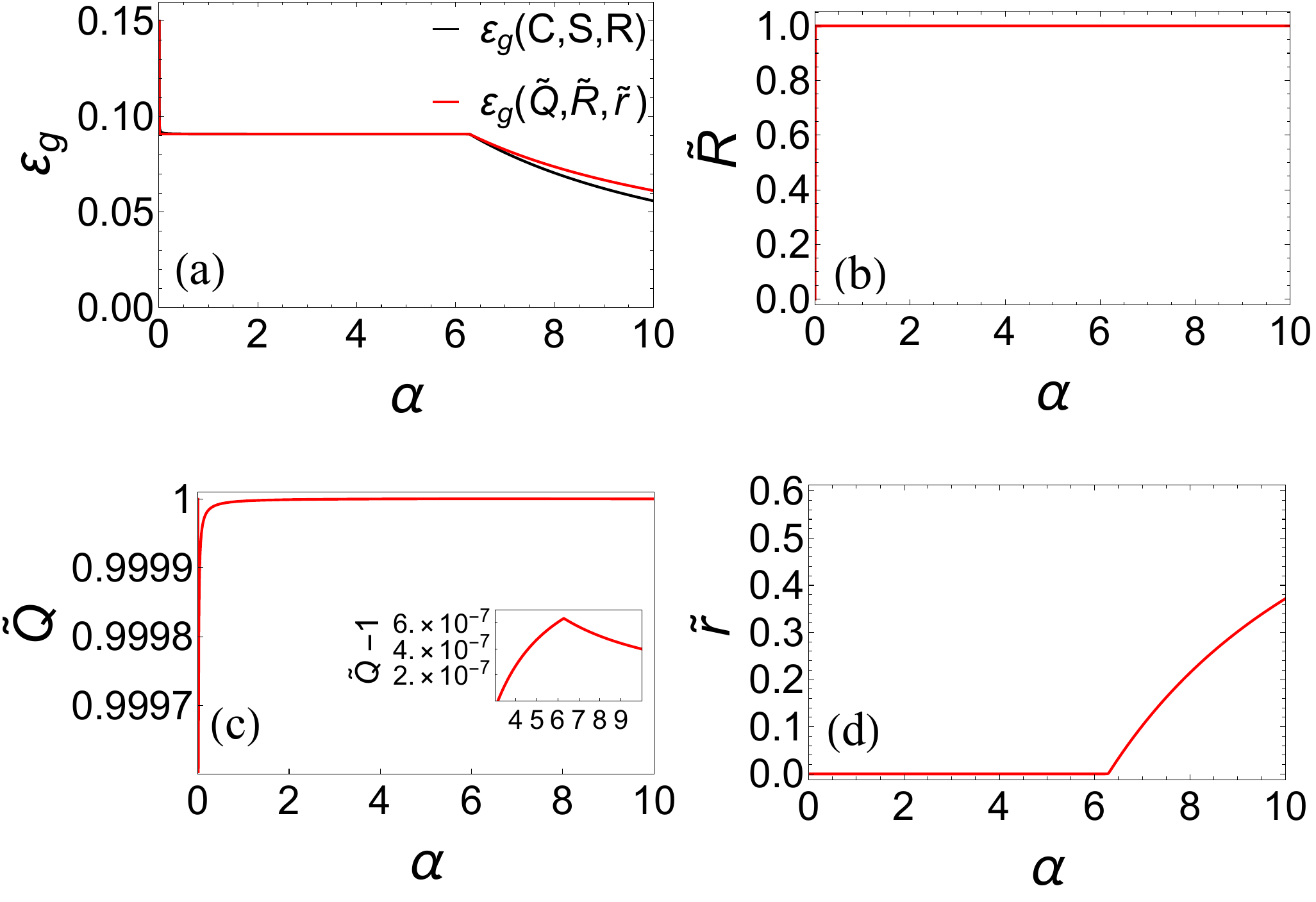}
    \caption{
    (a)~Generalization error for the realizable case $(K=M)$ with $(\gamma=10^{-6},K=10^{6})$, obtained by minimizing Eq.~(\ref{Eq:free_KeM}) and compared against results from \cite{Oostwal} in the $K\to\infty$ limit. Excellent agreement is observed for $\alpha<\alpha_{c}\approx2\pi$, while deviations appear deeper in the specialized phase ($\alpha>\alpha_{c}$) due to our expansion in order parameters.  (b)~$\tilde{R}$ remains $1$ for all $\alpha>0$.  
    (c)~$\tilde{Q}\approx 1$ in the unspecialized phase and near the transition, with a small correction $\mathcal{O}(1/K)$ (inset).
(d)~$\tilde{r}\sim\mathcal{O}(1/K)$ for $\alpha<\alpha_{c}$, then grows to $\tilde{r}\approx 1-2\pi/\alpha$ beyond $\alpha_{c}$. }
\label{Fig:Eg_KeM_inf}
\end{figure*}
%
\subsection{Solutions for $K=M$ with $\gamma \ll 1$}

When $K=M$, the student and teacher architectures match in complexity. This widely studied case was analyzed by Oostwal \emph{et~al.}\ \cite{Oostwal} for ReLU activations under $N\gg K$, which we reproduce for comparison. A common simplifying ansatz sets $Q_{ij}=\delta_{ij}+C(1-\delta_{ij})$ and $R_{ij}=R\,\delta_{ij}+S(1-\delta_{ij})$, leading to an unspecialized state ($R=S$) at low $\alpha$ and a continuous transition at $\alpha_{c}$ to a specialized state ($R>S$) as permutation symmetry among the student units is broken.

Within our formulation, the auxiliary variables in Eq.~(\ref{Eq:entropy1}) can be eliminated by neglecting terms of $\mathcal{O}(\gamma)$ for $\gamma\ll1$ while retaining contributions of order $\mathcal{O}(\gamma K)$ (see Appendix~A), yielding
\begin{align}
\nonumber f = & \alpha ~ K  \left[ \dfrac{\tilde{Q}}{8} -  \dfrac{\tilde{R}}{4} - \dfrac{\tilde{r}}{4 \pi} + \left(\dfrac{3}{8} - \dfrac{1}{2 \pi} \right) \right] - \frac{1}{2} \text{ln} \left[ \tilde{Q} - \tilde{R}^{2} \right]\\
\nonumber & - \dfrac{K(1- \gamma) -1}{2} ~ \text{ln} \left[ 1 - \tilde{r} - \dfrac{\tilde{Q} - \tilde{R}^{2}}{K}\right] \\
& - \dfrac{K \gamma}{2} \text{ln} \left[\tilde{r} - \dfrac{\tilde{R}^{2}}{K} \right] . \label{Eq:free_KeM} 
\end{align}
Technically, the learning curves can be obtained by solving the saddle-point equations for arbitrary $K$ with $\gamma \ll 1$. Given our expansion of Eq.~(\ref{Eq:Eg_expnd}), which neglects higher-order terms in $(Q_{ij},R_{ij})$, we expect quantitative accuracy in the unspecialized phase and near $\alpha_{c}$ where $R_{ij}=\mathcal{O}(1/K)$; deeper in the specialized regime, deviations from the exact learning curve arise as $R_{ii}$ grows with $\alpha$. Including higher-order terms in Eq.~(\ref{Eq:Eg_expnd}) would improve the description in that regime.

Figure~\ref{Fig:Eg_KeM10} shows numerical results for $(N=10^{12},\gamma=10^{-11}$, i.e., $K=10)$. In panel~(a), $\varepsilon_{g}(\tilde{Q},\tilde{R},\tilde{r})$ obtained using our formalism is compared with $\varepsilon_{g}(C,R,S)$ from Ref.~\cite{Oostwal}. The two agree in the unspecialized phase and near $\alpha_{c}\approx2\pi$, with differences appearing deeper in the specialized phase. Importantly, the qualitative phase structure is unchanged, and the two approaches agree asymptotically as $\alpha\to\infty$, where $\varepsilon_{g}\sim 1/\alpha$.

Panels~(b)--(d) of Fig.~\ref{Fig:Eg_KeM10} show the evolution of the order parameters $(\tilde{R},\tilde{Q},\tilde{r})$ with $\alpha$. In the unspecialized phase, permutation symmetry implies equal and small (of order $O(1/K)$) $R_{ij}$ and hence $\tilde{R}$ increases from zero but remains below unity; a kink at $\alpha_{c}$ marks the onset of specialization. In the specialized phase, eventually $R_{ii} \to 1$ and $R_{ij}\to 0$ ($i\neq j$), consistent with the order parameter value $\tilde{R} \equiv 1$ found everywhere in the specialized phase [panel~(b)]. Similarly, $\tilde{Q}$ rises to a maximum with a kink at $\alpha_{c}$ and then relaxes to $1$ for $\alpha>\alpha_{c}$ (each student unit aligns with a single normalized teacher unit). Finally, $\tilde{r}=\mathcal{O}(1/K)$ in the unspecialized phase -- consistent with small $O(1/K)$ and symmetric $R_{ij}$ -- and grows as $(\alpha-2\pi)/\alpha$ beyond $\alpha_{c}$.

For $K \gg 1$, assuming $\tilde{Q}$ and $\tilde{R}$ are $\mathcal{O}(1)$ while $\tilde{r}$ is $\mathcal{O}(1/K)$ in the unspecialized phase, a Taylor expansion gives
$\ln[1-\tilde{r}-(\tilde{Q}-\tilde{R}^{2})/K]\approx -(\tilde{r}+\tilde{Q}-\tilde{R}^{2}/K)$.
Solving the saddle-point equations yields
 \begin{subequations}
\begin{align}
\tilde{R} & = 1 - {1 \over K} 
\left({4 \pi - 2 \alpha \over \alpha \pi}\right)
+ \mathcal{O}(1/ K^2, \gamma)\\
\tilde{Q} & = 1 + { 1 \over K} \left({4 \alpha - 4 \pi \over \alpha \pi}\right) + \mathcal{O}(1/K^2, \gamma) \\
\tilde{r} & = 1/K + \mathcal{O}(1/K^2, \gamma) ~.
\end{align} 
\label{Eq:RQr_largeK}
\end{subequations}
The corresponding generalization error is 
 \begin{align}
    \varepsilon_{g} = \dfrac{1}{4} - \dfrac{1}{2 \pi}  +  \dfrac{1}{K} \left( \dfrac{1 }{2 \alpha} + \dfrac{1}{4\pi}\right) + \mathcal{O}(1/ K^2, \gamma) \ .
\end{align}
Thus a symmetric plateau at $\varepsilon_{g} \approx 0.09$ characterizes the unspecialized phase for $K \to \infty$ and $\gamma \ll 1$, while a correction of  $\mathcal{O}(1/K)$ describes a monotonically decreasing $\varepsilon_{g} $ for finite but large $K$.

On the other hand, in the specialized regime we find that $\tilde{r}$ is $\mathcal{O}(1)$, and the arguments of the second and third logarithms in Eq.~(\ref{Eq:free_KeM}) remain finite; the expansion used above is therefore not applicable. We find that the $\mathcal{O}(K\gamma)$ contributions to the entropy are negligible compared to other terms. Neglecting them, the saddle-point condition $\partial f/\partial \tilde{R}=0$ implies $\tilde{R}=1$ for all $\alpha$ (see Appendix~B), with
\begin{subequations}
\begin{align}
\tilde{Q} & = 1 + \dfrac{4 \pi}{\alpha \pi K + 2 \alpha} \\
\tilde{r} & = \dfrac{\alpha - 2 \pi}{\alpha} +  \dfrac{2\pi^{2}}{\alpha \pi K +2 \alpha} .
\end{align}
\end{subequations}
Substituting into Eq.~(\ref{Eq:Eg_QRr}) eliminates the $K$-dependence, and the generalization error leaves the symmetric plateau at $\alpha_{c}$ and decreases with $\alpha$ as
\begin{align}
\varepsilon_{g} = \left( \dfrac{1}{4} - \dfrac{1}{2 \pi} \right) - \dfrac{\alpha -2\pi }{4 \pi \alpha} . \label{Eq:Eg_spec}
\end{align} 
   
These results are confirmed numerically. Figure~\ref{Fig:Eg_KeM_inf} reports data for $( K = 10^{6}, \gamma = 10^{-6} )$. As in the finite-$K$ case, panel~(a) compares $\varepsilon_{g}(\tilde{Q},\tilde{R},\tilde{r})$ with the curve reproduced from Ref.~\cite{Oostwal}: excellent agreement is found for $\alpha < \alpha_{c}$. A continuous phase transition occurs at $\alpha_{c} \approx 2 \pi$,  followed by deviations for $\alpha > \alpha_{c}$ due to the order-parameter expansion. The order parameters behave as shown in panels (b)-(d): $\tilde{R}=1$ for all $\alpha>0$; $\tilde{Q}\approx 1+\mathcal{O}(1/K)$ near $\alpha_{c}$ with a kink at the transition (inset); and $\tilde{r} \approx (\alpha- 2\pi)/\alpha$ in the specialized regime. Note that, unlike in the symmetric phase, the behavior of $\varepsilon_{g}$ is independent of $K$ in the specialized phase, consistent with Fig.~\ref{Fig:KeqM_flatt}, where the learning curves for $K=10$ and $K=10^{6}$ coincide beyond $\alpha_{c}$.

\begin{figure*}[t!]
\centering
\includegraphics[width=15 cm, height=15 cm,keepaspectratio]{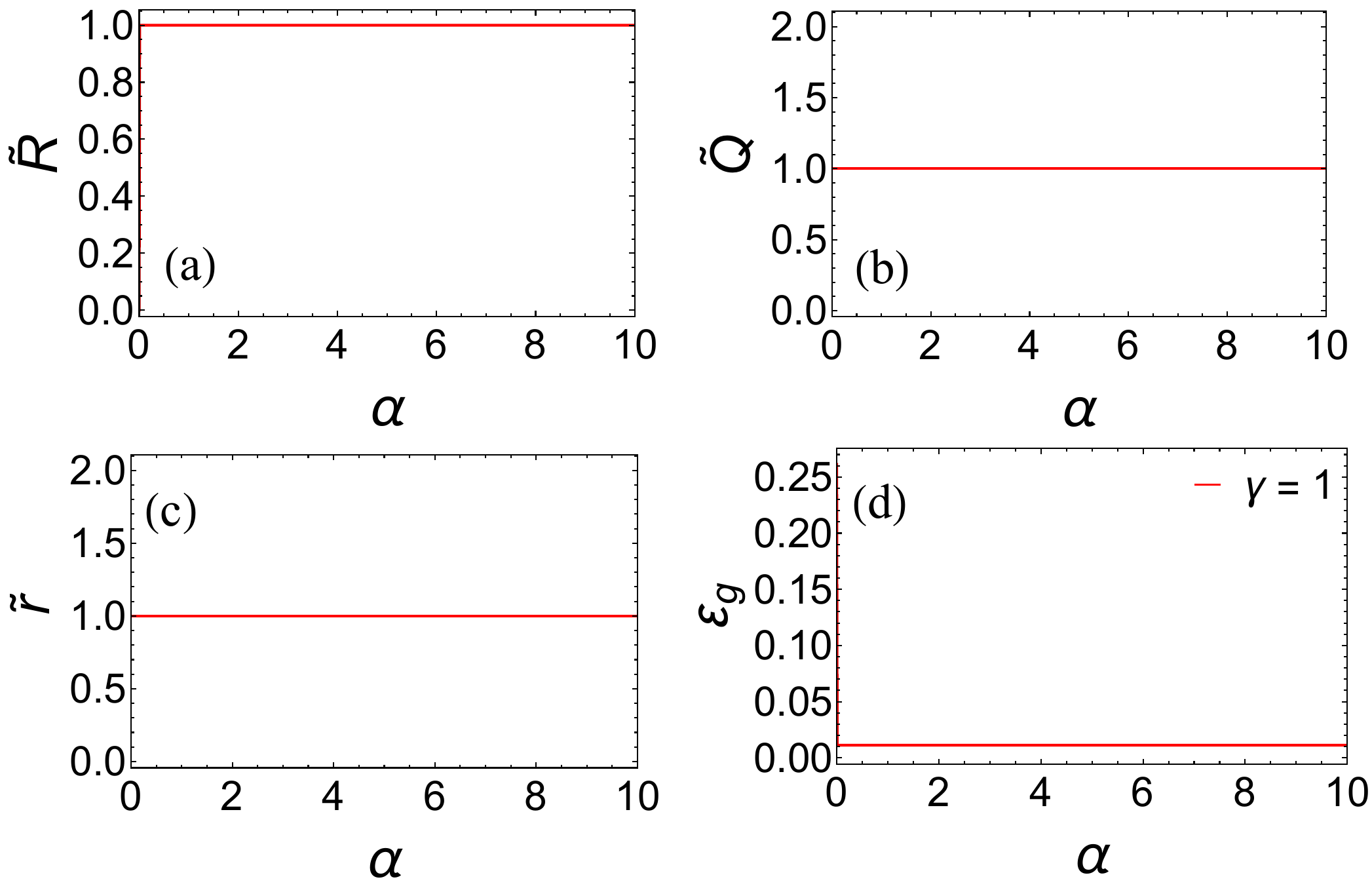}
\caption{ Learning curves for the realizable case $K=M=N=10^{12}$ with $\gamma=1$. The order parameters in panels (a–c), $\tilde{Q}$, $\tilde{R}$, and $\tilde{r}$, all remain equal to one, independent of $\alpha$.
(d)~The corresponding generalization error exhibits a constant plateau with height $\varepsilon_{g} \approx 0.01$.}
\label{Fig:Eg_QRr_1}
\end{figure*}

\subsection{Solutions for $K=M$ with $\gamma = 1$}

The case $\gamma = 1$ ($N=K=M$, with $N \gg 1$) exhibits no phase transition, as shown in Fig.~\ref{Fig:KeqM_flatt}. For $\gamma=1$, the entropic term Eq.~(\ref{Eq:entropy1}) simplifies greatly since all terms with prefactor $(1-\gamma)$ vanish. Minimizing the entropy with respect to the auxiliary variables shows that $\hat{\lambda}$ and $\hat{r}$ are coupled so that at the saddle point both $\tilde{r}$ and the norm ${\bm{J}}^{2}$ are constrained to one (Appendix~A).

 This peculiar constraint  $\tilde{r} = 1$ is compatible with two different limiting configurations of student weight vectors: the student vectors   are either in perfect alignment with the teacher vectors (specialized hidden units),  or there can be  completely random students (unspecialized hidden units). 
 In the first scenario, $R_{ii}\to 1$, $R_{ij}\to 0$ ($i\neq j$), and likewise for $Q_{ij}$, giving $\tilde{R}=\tilde{Q}=1$ and $\tilde{r}=1$. In the second scenario, for orthonormal teacher vectors and random student vectors with i.i.d.~Gaussian components of zero mean and unit variance, one finds
\begin{align}
\nonumber  \left\langle \tilde{r} \right\rangle &=  \dfrac{1}{N} \sum_{ij}^{N} \left\langle \left( \dfrac{{\bm{J}}_{i} \cdot {\bm{B}}_{j}}{N} \right)^{2} \right\rangle \\
\nonumber    &= \dfrac{1}{N} \sum_{ij}^{N} \sum_{lm}^{N} \dfrac{1}{N^{2}} \left\langle  J_{il} B_{jl} J_{im} B_{jm} \delta_{lm} \right\rangle \\
             &= 1 . \label{Eq:r_1}  
\end{align}
Moreover, $\tilde{Q} \approx 1$ since high-dimensional random vectors are nearly orthogonal (see the Model section), and consequently $\tilde{R} \approx 1$. Using that  $\tilde{r}=1$, one can eliminate the auxiliary variables in Eq.~(\ref{Eq:entropy1}), yielding the free energy
\begin{align}
\nonumber f =  & ~ \alpha ~ K  \left[ \dfrac{\tilde{Q}}{8} -  \dfrac{\tilde{R}}{4} - \dfrac{1}{4 \pi} + \left(\dfrac{3}{8} - \dfrac{1}{2 \pi} \right) \right] \\
 & - \dfrac{K-1}{2} \text{ln}\left[ 1- \dfrac{\tilde{Q}}{K} \right] - \frac{1}{2} \text{ln} \left[ \tilde{Q} - \tilde{R}^{2} \right] .  
\end{align}
For $\tilde{Q}=\mathcal{O}(1)$, the first logarithmic term can be expanded as
$\text{ln}[1-\tilde{Q}/K] \approx -\tilde{Q}/K$.
Solving the saddle-point equations in the large-$K$ limit gives $\tilde{R} = 1 - \mathcal{O}(1/K)$ and $\tilde{Q} = 1 + \mathcal{O}(1/K)$, which together with $\tilde{r}=1$ yields a plateau
\begin{align}
\varepsilon_{g} = \left(\dfrac{1}{4} - \dfrac{3}{4\pi} \right) + \mathcal{O}(1/ K) . 
\end{align}
Figure~\ref{Fig:Eg_QRr_1} shows results for $\gamma = 1$, $K = M = N = 10^{12}$: $\tilde{R}$, $\tilde{Q}$, and $\tilde{r}$ remain equal to one for all $\alpha$ [panels (a)- (c)], with a corresponding generalization-error plateau at $\varepsilon_{g} \approx 0.01$ [panel (d)], in agreement with the analytic prediction.

Our analysis demonstrates the absence of a phase transition in the SCM for finite $\gamma$, supporting the scenario of effectively random student vectors. Since for $K=N$ the student vectors still form a basis of the input space, a small generalization error is plausible without specialization. The alternative scenario, perfect learning (also compatible with $\tilde{r}=1$), is inconsistent with the observed nonzero plateau. Prior work has reported that the length of the symmetric plateau scales with learning rate and with the number of hidden units \cite{Richert_2022}. In the limit of an infinitely wide teacher, $M \to \infty$, the plateau is prolonged and the student remains in the symmetric phase, again consistent with the random-student scenario.

\subsection{Solutions in the asymptotic regime $\alpha \rightarrow \infty $}

To analyze the asymptotic behavior, recall that for large $\alpha$ one expects $R_{ii} \to 1$ and $R_{ij} \to 0$, consistent with an approach to perfect learning. This motivates the ansatz
$Q_{ij}=\delta_{ij} + (1- \delta_{ij}) q_{ij}$ and
$R_{ij} = (1- w_{ij}) \delta_{ij} + (1- \delta_{ij}) s_{ij}$,
with $q_{ij}, w_{ij}, s_{ij}$ small in the asymptotic regime. For $K = M$, rewriting Eq.~(\ref{Eq:Eg_ost}) in terms of these variables and expanding the nonlinear terms yields
\begin{align}
\nonumber 
\varepsilon_{g} = & \dfrac{1}{8 K} \sum_{i \neq j}^{K} q_{ij} + \dfrac{1}{2 K} \sum_{i}^{K} w_{i} - \dfrac{1}{4 K} \sum_{i \neq j}^{K} s_{ij} \\ 
                  & + \mathcal{O}(q_{ij}^{2},s_{ij}^{2},w_{i}^{3/2}).
\label{Eq:Eg_asump_expand}
\end{align}   
Analogously to the intermediate $\alpha$ case, we define aggregated order parameters 
\begin{align}
\nonumber \tilde{q} = \dfrac{1}{K} \sum_{i\neq j}^{K} & q_{ij} \quad,  \quad \tilde{w} = \dfrac{1}{K} \sum_{i=1}^{K} w_{i} \\
& \tilde{s} = \dfrac{1}{K} \sum_{i \neq j}^{K} s_{ij} ~ . \label{Eq:order2} 
\end{align}
Then  the generalization error Eq.~(\ref{Eq:Eg_asump_expand}) can be expressed in terms of the new order parameters as
\begin{align}
\varepsilon_{g} = \dfrac{\tilde{q}}{8} - \dfrac{\tilde{s}}{4} + \dfrac{\tilde{w}}{2} .
\label{Eq:Eg_asympt}
\end{align}
Next we compute the entropic part and find
\footnotesize
\begin{align}
\nonumber S &=  \dfrac{1}{N} \, \text{ln} \int \prod_{i=1}^{K} ~ d \mu(\underline{\bm{J}}_{i}) ~ \delta \left( \sum_{i \neq j}^{K}\bm{J}_{i} \cdot\bm{J}_{j} - N K \tilde{q} \right)\\
 & \times \delta \left(N K (1- \tilde{w}) - \sum_{i=1}^{K}\bm{J}_{i} \cdot\bm{B}_{i}\right) 
  \times \delta \left(N K \tilde{s} - \sum_{i \neq j}^{K}\bm{J}_{i} \cdot\bm{B}_{j} \right) \ .  \label{Eq:entrop2} 
\end{align} 
\normalsize
Similarly  to the previous case Eq.~(\ref{Eq:entrop}), the integral above can be evaluated using saddle point integration (see appendix.A), giving
\begin{align}
\nonumber S &= const. + \dfrac{1}{2} \text{ln} \left[1 + \tilde{q} - (1- \tilde{w} + \tilde{s})^{2} \right]  \\
  & + \dfrac{K-1}{2} \text{ln} \left[ 1- \dfrac{\tilde{q}}{K-1} - (1- \tilde{w} - \dfrac{\tilde{s}}{K-1})^{2} \right] .
  \label{Eq:entrop_asympt1}
\end{align}
Since no order parameter is defined as a sum over  higher powers of $(w_{i},s_{ij})$, the entropic term above has no $\gamma = M/N$ dependence. The free energy in the asymptotic regime is therefore
\begin{align}
\nonumber  f =& \alpha K \left[ \dfrac{\tilde{q}}{8} - \dfrac{\tilde{s}}{4} + \dfrac{\tilde{w}}{2} \right] - \dfrac{1}{2} \text{ln} \left[1 + \tilde{q} - (1- \tilde{w} + \tilde{s})^{2} \right]  \\
          & - \dfrac{K-1}{2} \text{ln} \left[ 1- \dfrac{\tilde{q}}{K-1} - (1- \tilde{w} - \dfrac{\tilde{s}}{K-1})^{2} \right] .
          \label{Eq:free_asymp}
\end{align}
Solutions to the saddle-point equations can be obtained numerically for arbitrary $K$, and analytically in both the large-$K$ limit and for a single hidden unit. Assuming $\tilde{q}, \tilde{w}, \tilde{s}$ are small as $\alpha\to\infty$, we expand the quadratic terms within the logarithms in Eq.~(\ref{Eq:entrop_asympt1}) and then neglect $\mathcal{O}(\tilde{w}^{2}, \tilde{s}^{2}, \tilde{w}\tilde{s})$ to obtain
\begin{align}
\nonumber S &= const. + \dfrac{1}{2} \text{ln}\left[ 2 \tilde{w} + \tilde{q} - 2 \tilde{s} \right] 
    + \dfrac{K-1}{2} \text{ln} \left[ 2 \tilde{w}- \dfrac{\tilde{q}- 2 \tilde{s}}{K-1}  \right] \\
 &= \dfrac{1}{2} \text{ln}\left[ 2 \tilde{w} + \tilde{q} - 2 \tilde{s} \right] + \dfrac{K-1}{2} \text{ln} \left[ \tilde{w} \right]  \ ,
\label{Eq:entrop_asympt2}
\end{align}
where the term $(\tilde{q}-2\tilde{s})/(K-1)$ was neglected in the last line for large $K$. The free energy reduces to
 \begin{align}
\nonumber  f =& \alpha K \left[ \dfrac{\tilde{q}}{8} - \dfrac{\tilde{s}}{4} + \dfrac{\tilde{w}}{2} \right] - \dfrac{1}{2} \text{ln}\left[ 2 \tilde{w} + \tilde{q} - 2 \tilde{s} \right] \\
  & - \dfrac{K-1}{2} \text{ln} \left[ \tilde{w} \right] + const.
\label{Eq:free_asymp_simplified}
\end{align}  
Solving the saddle point equations is now straightforward, and yields
\begin{subequations}
\begin{align}
2 \tilde{s} - \tilde{q} =& \dfrac{4}{\alpha} \left[ 1+ \mathcal{O}(1/K)\right] \\
\tilde{w} =& \dfrac{2}{\alpha} \left[ 1+ \mathcal{O}(1/K)\right] \ ,
\end{align}
\end{subequations}
and substitution into Eq.~(\ref{Eq:Eg_asympt}) gives the generalization error
\begin{align}
\varepsilon_{g} =  \dfrac{1}{2 \alpha}  \ .
\end{align}
For $K=1$ on the other hand, only one student-teacher overlap $\tilde{w}$ is required, and the free energy Eq.~(\ref{Eq:free_asymp_simplified}) simplifies greatly to become
\begin{align}
    f = \alpha \dfrac{\tilde{w}}{2} - \dfrac{1}{2} \text{ln}\left[ 2 \tilde{w} \right]
\end{align}
Minimization gives $\tilde{w}=1/\alpha$ and $\varepsilon_{g}=1/(2\alpha)$, in agreement with \cite{Oostwal}. Remarkably, the asymptotic learning behavior of the SCM coincides for finite $K$ and for $K \to \infty$. The same asymptotic scaling has been reported previously for SCMs with alternative activation functions, notably the error function activation \cite{Oostwal,Biehl_1998}.

\section{Conclusion}

We have analyzed the behavior of the soft committee machine (SCM) with ReLU activation within the annealed approximation, using a statistical mechanics formulation of the student-teacher scenario. Across different learning regimes -- ranging from the standard case $N \gg K$ to the ultra-wide regime $K \ge N$ -- the model exhibits qualitatively similar behavior as long as the number of teacher hidden units satisfies $M \ll N$. In this regime, learning proceeds through a continuous transition from an unspecialized state, where the student's hidden units remain permutation symmetric, to a specialized state, in which each student unit learns a distinct teacher rule.

This phase transition is second order and occurs at a critical data load $\alpha_{c} \approx 2\pi$ for small $\gamma = M/N$. Our formulation reproduces the well-established results for SCMs with ReLU activations \cite{Oostwal}, confirming that for $\gamma \ll 1$ the generalization error $\varepsilon_{g}$ displays a distinct symmetric plateau followed by a transition to a specialized phase. For finite $\gamma$, however, the transition disappears: $\varepsilon_{g}$ decreases smoothly with $\alpha$, and for $\gamma = 1$ the system remains on a low plateau independent of $\alpha$. These results emphasize the crucial role of the network dimensions $(N, K, M)$ in determining learning dynamics, and demonstrate that conventional mean-field analyses must be reconsidered in ultra-wide architectures.

Modern machine learning often invokes the "double descent" phenomenon \cite{doubledecent} to explain the success of over-parameterized models, whose behavior can be linked to Gaussian processes and neural tangent kernels (NTKs) \cite{Jacot2018,lee2018,Arora2019}. Our results, however, do not show enhanced generalization in the ultra-wide limit beyond a modest reduction in the plateau height. This suggests that the statistical mechanics picture of the SCM, even when extended to $K \ge N$, remains qualitatively distinct from the NTK regime, expected for $K\to \infty $ and a finite input dimension $N$.

We also find that in the asymptotic limit $\alpha \to \infty$, the generalization error scales as $\varepsilon_{g} \propto 1/\alpha$, independent of $\gamma$ and $K$. This asymptotic form coincides with earlier results for SCMs employing other activation functions \cite{Oostwal,Biehl_1998}, indicating that our framework captures universal features of the high-data regime.
 
Finally, our formulation -- based on the aggregated order parameters $(\tilde{Q}, \tilde{R}, \tilde{r})$ -- provides a unified description valid across the full range of $(N,K,M)$ and can be readily generalized to other activation functions, provided that the expansion of the nonlinear terms in the generalization error remains controlled. Extending this approach to compute the quenched free energy, using the replica method, would allow one to incorporate finite-temperature effects and fluctuations beyond the annealed approximation, offering a deeper statistical mechanics understanding of learning in shallow networks.

Note added: After completion of our work we became aware of related work in Ref.~\cite{Barbier25}, which studies feature learning of a multi-layer perceptron whose width scales like the input dimension. 
 
\begin{acknowledgments} 
 We thank the Center for Scalable Data Analytics and Artificial Intelligence (Scads.AI), Dresden/Leipzig, for their support with funding and computation resources. 
\end{acknowledgments}


\appendix

\section{Derivation of the entropic term for different scenarios }  
To compute the entropic term for general $(K,M,N)$, we start from the definition of the entropic term
\footnotesize
\begin{align}
\nonumber S = \dfrac{1}{N} \, \text{ln} & \int  \prod_{i=1}^{K}  \left(  \dfrac{d\bm{J}_{i}}{(2 \pi e)^{N/2}} \delta(N-\bm{J}^{2}_{i}) \right) ~  \delta \left( \sum_{ij=1}^{K}\bm{J}_{i} \cdot\bm{J}_{j} - \dfrac{N K^{2}}{M} \tilde{Q} \right) \times \\
& \, \delta \left(N K \tilde{R} - \sum_{i=1}^{K} \sum_{j=1}^{M}\bm{J}_{i} \cdot\bm{B}_{j}\right) \, \delta \left(N^{2} K \tilde{r} - \sum_{i=1}^{K} \sum_{j=1}^{M} (\bm{J}_{i} \cdot\bm{B}_{j})^{2} \right)  .
\end{align}
\normalsize
Next, we introduce the integral representation of the delta function
\begin{align}
\delta(x-a) = \int_{-i \infty}^{i \infty} ~ \dfrac{d \hat{x}}{2 i \pi} ~ e^{\hat{x}(x-a)} ,  \label{Eq:delta func} 
\end{align}
we use $\hat{Q}, \hat{R}, \hat{r}$ as the auxiliary variables of the order parameters $\tilde{Q}, \tilde{R}$ and $\tilde{r}$ respectively, in addition to $\hat{\lambda}$ for the normalization condition, one obtain 
\begin{align}
\nonumber S = & K \hat{\lambda} + K \hat{R} \tilde{R} + N K \hat{r} \tilde{r} - \dfrac{K^{2}}{M} \hat{Q} \tilde{Q} \\
\nonumber & + \dfrac{1}{N} \text{ln} \int ~ \prod_{i=1}^{K} \dfrac{d\bm{J}_{i}}{(2 \pi e)^{N/2}}~ \text{exp} \left( - \hat{\lambda}  \sum_{i=1}^{K}\bm{J}^{2}_{i} + \hat{Q} \sum_{ij=1}^{K}\bm{J}_{i} \cdot\bm{J}_{j} \right. \\
   & \left. - \hat{R} \sum_{i=1}^{K} \sum_{j=1}^{M}\bm{J}_{i} \cdot\bm{B}_{j} - \hat{r} \sum_{i=1}^{K} \sum_{j=1}^{M} (\bm{J}_{i} \cdot\bm{B}_{j})^{2}
              \right) 
\label{Eq:S_initial}
\end{align}
\small
now we define the vectors
\begin{align}
\tilde{\bm{J}}^{(N K \times 1 )} = 
\begin{pmatrix}
\bm{J}_{1}  \\
\bm{J}_{2} \\
\vdots \\
\bm{J}_{K}
\end{pmatrix}, \quad
\bm{\mathcal{B}}^{(N K \times 1 )} = 
\begin{pmatrix}
\overline{\bm{B}}  \\
\overline{\bm{B}} \\
\vdots \\
\overline{\bm{B}}
\end{pmatrix}, \quad \text{with} ~ \overline{\bm{B}} = \sum_{j=1}^{M}\bm{B}_{j}
\end{align}
\normalsize
which allow us to rewrite the integral over the student and teacher wight vectors in a Gaussian form with the $(K \times K)$ block matrix 
\begin{align*}
A^{(K \times K)} = D^{N \times N} \delta_{ij} + O^{N \times N} (1-\delta_{ij})                            
\end{align*}
where the diagonal and off-diagonal block matrices elements are   
\begin{align*} 
 D_{l m} = (2 \hat{\lambda}- 2 \hat{Q} + 2 N \hat{r}_{l m \le M}) \delta_{l m} \quad \text{and} \quad O_{l m} = - 2 \hat{Q}  \delta_{l m} ~ ,
\end{align*}
here, the auxiliary variable $\hat{r}$ is scaled with $N$ due to the orthonormal choice of the teacher wight vectors. Thus, Eq.~(\ref{Eq:S_initial}) now reads  
\begin{align}
\nonumber & S  =   K \hat{\lambda} + K \hat{R} \tilde{R} + N K \hat{r} \tilde{r} - \dfrac{K^{2}}{M} \hat{Q} \tilde{Q} \\ & + \dfrac{1}{N} \text{ln} \int ~ \prod_{i=1}^{K} \dfrac{d\bm{J}_{i}}{(2 \pi e)^{N/2}}~ \text{exp} \left[ -\dfrac{1}{2} \left( \tilde{\bm{J}}^{T} A \tilde{\bm{J}} + 2 \hat{R}~ \bm{\mathcal{B}}^{T} \tilde{\bm{J}} \right) \right] .
\end{align}
It is straight forward to compute the Gaussian integral, one obtain
\begin{align}
\nonumber S = \min_{\substack{\hat{\lambda}, \hat{Q} \\ \hat{R},\hat{r}}} & \left\lbrace -\dfrac{K}{2} + K \hat{\lambda} + K \hat{R} \tilde{R} + N K \hat{r} \tilde{r} - \dfrac{K^{2}}{M} \hat{Q} \tilde{Q} \right. \\ & \left.  - \dfrac{1}{2 N} \text{ln} \det A + \dfrac{1}{2 N}~ \hat{R}^{2} \bm{\mathcal{B}}^{T}  A^{-1} \bm{\mathcal{B}}
\right\rbrace 
\end{align}   
Diagonalizing the symmetric matrix $A$, then the determinant of the matrix can be computed as the product of its eigenvalues. The degeneracy of the each eigenvalue depends explicitly on the choice of $N, M$ and $K$. One obtain the entropy in terms of the order parameters as
\begin{align}
\nonumber S = \min_{\substack{\hat{\lambda}, \hat{Q} \\ \hat{R},\hat{r}}} & \left\lbrace -\dfrac{K}{2} - \dfrac{K}{2} \text{ln}2 + K \hat{\lambda} + K \hat{R} \tilde{R} + N K \hat{r} \tilde{r} - \dfrac{K^{2}}{M} \hat{Q} \tilde{Q} 
\right. \\ \nonumber & \left.  - \dfrac{(N-M)(K-1)}{2 N} \text{ln} \hat{\lambda} - \dfrac{M(K-1)}{2 N} \text{ln} (\hat{\lambda} + N \hat{r})
 \right. \\ \nonumber & \left. - \dfrac{(N-M)}{2 N} \text{ln} (\hat{\lambda} - K \hat{Q}) - \dfrac{M}{2 N} \text{ln} (\hat{\lambda} + N \hat{r} - K \hat{Q}) \right. \\ & \left. + \dfrac{\hat{R}^{2}}{4} ~ \dfrac{K M}{(\hat{\lambda} + N \hat{r} - K \hat{Q})}
\right\rbrace 
\end{align}  
finally define the ratio $\gamma = M/N$ and rescale the variable $N \hat{r} \to \hat{r} $, yields Eq.(\ref{Eq:entropy1}).

\subsection{Derivation of the entropic term for $\gamma \ll 1$ }
For $\gamma \ll 1$, terms of order $\gamma$ contribution to the entropy is very small compared to other terms and can be neglected. So the entropy is given by
\begin{align}
\nonumber S = \min_{\substack{\hat{\lambda}, \hat{Q} \\ \hat{R},\hat{r}}} & \left\lbrace -\dfrac{K}{2} - \dfrac{K}{2} \text{ln}2 + K \hat{\lambda} + K \hat{R} \tilde{R} + K \hat{r} \tilde{r} 
\right. \\ \nonumber & \left. - \dfrac{K^{2}}{M} \hat{Q} \tilde{Q}  - \dfrac{K(1- \gamma)-1}{2} \text{ln} \hat{\lambda} - \dfrac{\gamma K}{2} \text{ln} (\hat{\lambda} + \hat{r})
 \right. \\  & \left. - \dfrac{1}{2} \text{ln} (\hat{\lambda} - K \hat{Q}) + \dfrac{\hat{R}^{2}}{4} ~ \dfrac{K M}{(\hat{\lambda} +  \hat{r} - K \hat{Q})}  + \mathcal{O}(\gamma)
\right\rbrace \label{Eq:S_gamm_small} 
\end{align} 
Solving the saddle point equations  
$$ \dfrac{\partial S}{\partial \hat{R}} ~ = ~ \dfrac{\partial S}{\partial \hat{Q}} ~ = ~ \dfrac{\partial S}{\partial \hat{r}} ~ = ~ \dfrac{\partial S}{\partial \hat{\lambda}} ~ = ~ 0  $$  
yields at the saddle point :
\begin{subequations}
\begin{align}
 \hat{R} = & -\frac{2}{M} ~ \tilde{R} (\hat{\lambda} +  \hat{r} - K \hat{Q}) \\
 K \hat{Q} = & \hat{\lambda} - \dfrac{M}{2 K (\tilde{Q} - \tilde{R}^{2})} \\
\hat{r} = & \dfrac{\gamma}{2} \dfrac{1}{\tilde{r} - \tilde{R}^{2}/M} - \hat{\lambda} \\
\hat{\lambda} = & \dfrac{K(1-\gamma)-1}{2 K} ~ \dfrac{1}{1- \tilde{r} - (\tilde{Q} - \tilde{R}^{2})/M}
\end{align}
\end{subequations}
Now we substitute these solutions back into Eq.(\ref{Eq:S_gamm_small}) then after some algebra one obtain
\begin{align}
\nonumber S = & const. +  \dfrac{K(1- \gamma) -1}{2} ~ \text{ln} \left[ 1 - \tilde{r} - \dfrac{\tilde{Q} - \tilde{R}^{2}}{M}\right] \\
&+ \dfrac{K \gamma}{2} \text{ln} \left[\tilde{r} - \dfrac{\tilde{R}^{2}}{M} \right] + \frac{1}{2} \text{ln} \left[ \tilde{Q} - \tilde{R}^{2} \right] .
\end{align} 
\subsection{Derivation of the entropic term for $\gamma = 1$ }

For $\gamma = 1$, all terms with prefactor $(1-\gamma)$ in Eq.~(\ref{Eq:entropy1}) vanishes leading to
\begin{align}
\nonumber S = \min_{\substack{\hat{\lambda}, \hat{Q} \\ \hat{R},\hat{r}}} & \left\lbrace -\dfrac{K}{2} - \dfrac{K}{2} \text{ln}2 + K \hat{\lambda} + K \hat{R} \tilde{R} + K \hat{r} \tilde{r} - \dfrac{K^{2}}{M} \hat{Q} \tilde{Q} 
\right. \\  
\nonumber & \left.  - \dfrac{K-1}{2} \text{ln} (\hat{\lambda} + \hat{r})
  - \dfrac{1}{2} \text{ln} (\hat{\lambda}+ \hat{r} - K \hat{Q}) \right. \\
   & \left. + \dfrac{\hat{R}^{2}}{4} ~ \dfrac{K M}{(\hat{\lambda} +  \hat{r} - K \hat{Q})}
\right\rbrace  \label{Eq:S_gamm_1} 
\end{align}
One need to solve the saddle point equations :
\small
\begin{align}
 & K \tilde{R} + \dfrac{\hat{R}}{2} ~ \dfrac{K M}{(\hat{\lambda} +  \hat{r} - K \hat{Q})} = 0 \label{Req} \\
 &-\dfrac{K^{2}}{M} \tilde{Q} + \frac{K}{2} \dfrac{1}{\hat{\lambda} +\hat{r}- K \hat{Q}} + \dfrac{\hat{R}^{2}}{4} ~ \dfrac{K^{2} M}{(\hat{\lambda} +  \hat{r} - K \hat{Q})^{2}} = 0 \label{Qeq}\\
& K \tilde{r} - \frac{K-1}{2} \dfrac{1}{\hat{\lambda} + \hat{r}} - \frac{1}{2} \dfrac{1}{\hat{\lambda} +\hat{r}- K \hat{Q}}- \dfrac{\hat{R}^{2}}{4} ~ \dfrac{K M}{(\hat{\lambda} +  \hat{r} - K \hat{Q})^{2}} = 0 \label{req}\\
& K - \frac{K-1}{2} \dfrac{1}{\hat{\lambda} + \hat{r}} - \frac{1}{2} \dfrac{1}{\hat{\lambda} +\hat{r}- K \hat{Q}} - \dfrac{\hat{R}^{2}}{4} ~ \dfrac{K M}{(\hat{\lambda} +  \hat{r} - K \hat{Q})^{2}} = 0 \label{lamdaeq}
\end{align}
\normalsize
From \ref{Req} we obtain 
\begin{align}
\hat{R} &= -\frac{2}{M} ~ \tilde{R} (\hat{\lambda} +  \hat{r} - K \hat{Q}) \label{Eq:Rhat1} 
\end{align}
substitute $\hat{R}$ in \ref{Qeq},
\begin{align}
\frac{1}{2} \dfrac{1}{\hat{\lambda} +\hat{r} -K \hat{Q}} = \dfrac{K}{M} (\tilde{Q} - \tilde{R}^{2})  \label{Eq:Qhat1} 
\end{align}
next, substitute \ref{Eq:Rhat1} and \ref{Eq:Qhat1} in \ref{req}
\begin{align}
\frac{K-1}{2} \dfrac{1}{\hat{\lambda} + \hat{r}} = K (\tilde{r} - \tilde{Q}/M) \label{Eq:rhat1} 
\end{align}
Substituting \ref{Eq:Rhat1},\ref{Eq:Qhat1} and \ref{Eq:rhat1} in \ref{lamdaeq}, one found that $\tilde{r} = 1$ at the saddle point which implies that $\hat{\lambda}$ and $\hat{r}$ are coupled together. Thus, solutions to the saddle point equations are 
\begin{subequations}
\begin{align}
 \hat{R} = & -\frac{2}{M} ~ \tilde{R} (\hat{\lambda} + \hat{r} - K \hat{Q}) \\
  K \hat{Q} = & \hat{\lambda}+\hat{r} - \dfrac{M}{2 K (\tilde{Q} - \tilde{R}^{2})} \\
 \hat{\lambda} + \hat{r} = & \dfrac{K-1}{2 K} \dfrac{1}{1 - \tilde{Q}/M}
\end{align}
\end{subequations}
Finally, eliminating the auxiliary variables in \ref{Eq:S_gamm_1} one obtain
\begin{align}
S = const. + \dfrac{K-1}{2} \text{ln} \left[1 - \dfrac{\tilde{Q}}{M} \right] + \frac{1}{2} \text{ln} \left[ \tilde{Q} - \tilde{R}^{2} \right] 
\end{align}

\subsection{Derivation of the entropic term in the asymptotic regime $\alpha \rightarrow \infty$ }
Here we start from Eq.~(\ref{Eq:entrop2}) then using the integral representation of the delta function Eq.~(\ref{Eq:delta func}), one obtain
\begin{align}
\nonumber S & =   K \hat{\lambda} + K \hat{w} (1- \tilde{w}) +  K \hat{s} \tilde{s} - K \hat{q} (1+\tilde{q}) \\
\nonumber &+ \dfrac{1}{N} \text{ln} \int ~ \prod_{i=1}^{K} \dfrac{d\bm{J}_{i}}{(2 \pi e)^{N/2}} \\ 
& \times \text{exp} \left[ -\dfrac{1}{2} \left( \tilde{\bm{J}}^{T} A \tilde{\bm{J}} + 2 \left( (\hat{w} - \hat{s} ) ~ \tilde{\bm{B}} + \hat{s} ~ \bm{\mathcal{B}} \right)^{T} \cdot \tilde{\bm{J}} \right) \right] .
\end{align} 
with 
\small
\begin{align}
\tilde{\bm{J}}^{(N K \times 1 )} = 
\begin{pmatrix}
\bm{J}_{1}  \\
\bm{J}_{2} \\
\vdots \\
\bm{J}_{K}
\end{pmatrix},
\tilde{\bm{B}}^{(N K \times 1 )} = 
\begin{pmatrix}
\bm{B}_{1}  \\
\bm{B}_{2} \\
\vdots \\
\bm{B}_{K}
\end{pmatrix},
\bm{\mathcal{B}}^{(N K \times 1 )} = 
\begin{pmatrix}
\overline{\bm{B}}  \\
\overline{\bm{B}} \\
\vdots \\
\overline{\bm{B}}
\end{pmatrix} 
\end{align}
\normalsize
with $\overline{\bm{B}} = \sum_{j=1}^{K}\bm{B}_{j}$ and the $(K \times K)$ block matrix 
\begin{align*}
A^{(K \times K)} = (2 \hat{\lambda} - 2 \hat{q})~ I^{N \times N} \delta_{ij} + (-2 \hat{q}) ~ I^{N \times N} (1-\delta_{ij})                             
\end{align*}
where $I$ denotes an $(N \times N)$ unit matrix. Evaluating the Gaussian integral yields
\begin{align}
\nonumber S = \min_{\substack{\hat{\lambda}, \hat{q} \\ \hat{w},\hat{s}}} & \left\lbrace -\dfrac{K}{2}+ K \hat{\lambda} + K \hat{w}( 1 - \tilde{w}) +  K \hat{s} \tilde{s} \right. \\
\nonumber & \left. - K \hat{q} ( 1+ \tilde{q}) - \dfrac{1}{2 N} \text{ln} \det A \right. \\
& \left.+ \dfrac{1}{2 N}~ \sum_{ij}^{K} \left(  (\hat{w}- \hat{s}) \bm{B}_{i} + \hat{s} \,\overline{\bm{B}} \right)^{T}  A^{-1}_{ij} \left(  (\hat{w}- \hat{s}) \bm{B}_{j} + \hat{s}\,  \overline{\bm{B}} \right)
\right\rbrace  
\end{align}
Next we compute the determinant of $A$ and the sum over the elements of the inverse matrix which give 
\begin{align}
\nonumber S = \min_{\substack{\hat{\lambda}, \hat{q} \\ \hat{w},\hat{s}}} & \left\lbrace -\dfrac{K}{2} -\dfrac{K}{2} \text{ln} 2 + K \hat{\lambda} + K \hat{w}( 1 - \tilde{w}) +  K \hat{s} \tilde{s} - K \hat{q} ( 1+ \tilde{q})  \right. \\ 
\nonumber & \left.  - \dfrac{K-1}{2} \text{ln}(\hat{\lambda}) - \dfrac{1}{2} \text{ln}(\hat{\lambda} - K \hat{q}) + \dfrac{K (\hat{w}- \hat{s})^{2}}{4} \, \dfrac{\hat{\lambda} - (K-1) \hat{q}}{\hat{\lambda}(\hat{\lambda} - K \hat{q})} \right. \\
& \left. + \dfrac{K}{4} \, \dfrac{2 (\hat{w}- \hat{s}) \hat{s} + K \hat{s}^{2}}{(\hat{\lambda} - K \hat{q})}
\right\rbrace  .
\end{align} 
To facilitate the calculations of the saddle point solutions we introduce a new auxiliary variable $\hat{\Delta} = \hat{w} - \hat{s}$, then rewrite the entropic term as  
\begin{align}
\nonumber S = \min_{\substack{\hat{\lambda}, \hat{q} \\ \hat{\Delta},\hat{s}}} & \left\lbrace -\dfrac{K}{2} -\dfrac{K}{2} \text{ln} 2 + K \hat{\lambda} + K \hat{\Delta}( 1 - \tilde{w}) +  K \hat{s} ( ( 1 - \tilde{w}) + \tilde{s}) \right. \\
\nonumber & \left. - K \hat{q} ( 1+ \tilde{q}) - \dfrac{K-1}{2} \text{ln}(\hat{\lambda})  - \dfrac{1}{2} \text{ln}(\hat{\lambda} - K \hat{q}) \right. \\
& \left. + \dfrac{K \hat{\Delta}^{2}}{4} \, \dfrac{\hat{\lambda} - (K-1) \hat{q}}{\hat{\lambda}(\hat{\lambda} - K \hat{q})} + \dfrac{K}{4} \, \dfrac{2  \hat{\Delta} \hat{s} + K \hat{s}^{2}}{(\hat{\lambda} - K \hat{q})} 
\right\rbrace  .  \label{Eq:entropy_asymp} 
\end{align}
Solving the saddle point equations give the auxiliary variables as a function of $\tilde{r}$, $\tilde{s}$ and $\tilde{w}$, one obtain  
\begin{align}
K \hat{s} = & \hat{\Delta} + 2 \left( ( 1 - \tilde{w}) + \tilde{s}\right) \left( K \hat{q} - \hat{\lambda}\right) \\
(K-1) \hat{\Delta} = & 2 \hat{\lambda} \left( \tilde{s} - (K-1)( 1 - \tilde{w}) \right) \\
K \hat{q} = & \hat{\lambda} + \dfrac{1}{2\left( 1+( 1 - \tilde{w})+\tilde{s} \right) \left( ( 1 - \tilde{w}) + \tilde{s} -1\right) -2 \tilde{q} } \\
\dfrac{1}{2 \hat{\lambda}} = & \left( 1- \dfrac{\tilde{q}}{K-1} - \left( ( 1 - \tilde{w}) - \dfrac{\tilde{s}}{K-1} \right)^{2}  \right)    
\end{align}
Substitute these solutions back into Eq.~(\ref{Eq:entropy_asymp}) yields finally the entropic term 
\begin{align}
\nonumber S &= const. + \dfrac{1}{2} \text{ln} \left[1 + \tilde{q} - (1- \tilde{w} + \tilde{s})^{2} \right]  \\
            & + \dfrac{K-1}{2} \text{ln} \left[ 1- \dfrac{\tilde{q}}{K-1} - (1- \tilde{w} - \dfrac{\tilde{s}}{K-1})^{2} \right] 
\end{align}

\renewcommand{\theequation}{B\arabic{equation}}
\setcounter{equation}{0}
\section{The realizable case $K=M, \gamma \ll 1$ saddle point calculations }
When starting from the free energy Eq.~(\ref{Eq:free_KeM}), one obtains two distinct sets of solutions.

\subsection{The unspecialized phase solutions } 

Here we have $\tilde{r}$ of order $\mathcal{O}(1/K)$ while $\tilde{Q}$ and $\tilde{R}$ are of order one, so one can expand the logarithmic term 
 $\text{ln} [1- \tilde{r} - (\tilde{Q} - \tilde{R}^{2})/K] \approx - (\tilde{r}+ \tilde{Q} - \tilde{R}^{2}/K ) $ , hence, the saddle point equations are : 
\begin{align}
 \dfrac{\alpha K}{4} &+ \dfrac{K (1- \gamma)-1}{K} - \dfrac{1}{\tilde{Q}- \tilde{R}^{2}} = 0 \label{fQ0eq} \\
  -\dfrac{\alpha K}{4 \pi} &+ \dfrac{K (1- \gamma)-1}{2} - \dfrac{K \gamma}{2} \dfrac{1}{\tilde{r} - \tilde{R}^{2}/K} = 0 \label{fr0eq} \\
 -\dfrac{\alpha K}{4} &- \dfrac{K (1- \gamma)-1}{K} \tilde{R} + \dfrac{\tilde{R}}{\tilde{Q}- \tilde{R}^{2}} + \dfrac{\gamma \tilde{R}}{\tilde{r} - \tilde{R}^{2}/K} = 0 \label{fR0eq}
\end{align}
From \ref{fQ0eq} and \ref{fr0eq} we have
\begin{align}
\dfrac{1}{\tilde{Q}- \tilde{R}^{2}} &= \dfrac{\alpha K}{4} + \dfrac{K (1- \gamma)-1}{K} \label{fQ0eq2} \\
\dfrac{\gamma}{\tilde{r} - \tilde{R}^{2}/K} &= -\dfrac{\alpha}{2 \pi} + \dfrac{K (1- \gamma)-1}{K} \label{fr0eq2}
\end{align}
substitute \ref{fQ0eq2} and \ref{fr0eq2} in \ref{fR0eq}, one finds  
\begin{align}
 \tilde{R} = \dfrac{\alpha \pi K^{2}}{\alpha (\pi K^{2} - 2 K) + 4 \pi (K(1-\gamma)-1)}
\end{align}
Substitute $\tilde{R}$ back into \ref{fQ0eq} and \ref{fr0eq}, one obtain 
\begin{align}
 \tilde{Q} &= \tilde{R}^{2} + \dfrac{4 K}{\alpha K^{2} + 4K(1-\gamma) -4} \\
 \tilde{r} &= \dfrac{\tilde{R}^{2}}{K} + \dfrac{2 \pi K \gamma}{2 \pi K (1- \gamma) - 2 \pi - \alpha K} ,
\end{align} 
which for large $K$ and $\gamma \ll 1$ yields Eq.~(\ref{Eq:RQr_largeK}) . \\ 
%
\subsection{The specialized phase solutions}

Here, terms of order $\mathcal{O}(K \gamma)$ are negligible in comparison to the other terms in the entropic part, the free energy now reads 
\begin{align}
\nonumber f = & \alpha ~ K  \left[ \dfrac{\tilde{Q}}{8} -  \dfrac{\tilde{R}}{4} - \dfrac{\tilde{r}}{4 \pi} + \left(\dfrac{3}{8} - \dfrac{1}{2 \pi} \right) \right] - \frac{1}{2} \text{ln} \left[ \tilde{Q} - \tilde{R}^{2} \right]\\
 & - \dfrac{K -1}{2} ~ \text{ln} \left[ 1 - \tilde{r} - \dfrac{\tilde{Q} - \tilde{R}^{2}}{K}\right] .
\end{align}
minimizing the free energy with respect to $\tilde{Q},\tilde{R}$ and $\tilde{r}$ give:
\begin{align}
 \dfrac{\alpha K}{4} + \dfrac{K-1}{K} ~ \dfrac{1}{1- \tilde{r} - \dfrac{\tilde{Q} - \tilde{R}^{2}}{K}} - \dfrac{1}{\tilde{Q}- \tilde{R}^{2}}  &= 0 \label{fQeq} \\
  -\dfrac{\alpha }{2 \pi} + \dfrac{K-1}{ K} ~ \dfrac{1}{1- \tilde{r} - \dfrac{\tilde{Q} - \tilde{R}^{2}}{K}}   &= 0 \label{freq} \\
 -\dfrac{\alpha K}{4} - \dfrac{K-1}{K} ~ \dfrac{\tilde{R}}{1- \tilde{r} - \dfrac{\tilde{Q} - \tilde{R}^{2}}{K}} + \dfrac{\tilde{R}}{\tilde{Q}- \tilde{R}^{2}} &= 0 . \label{fReq}
\end{align}
from \ref{fQeq} and \ref{freq} we have 
\begin{align}
\dfrac{\alpha K}{4} + \dfrac{K-1}{K} ~ \dfrac{1}{1- \tilde{r} - \dfrac{\tilde{Q} - \tilde{R}^{2}}{K}} &= \dfrac{1}{\tilde{Q}- \tilde{R}^{2}}   \label{I} \\
\dfrac{1}{1- \tilde{r} - \dfrac{\tilde{Q} - \tilde{R}^{2}}{K}}  &= \dfrac{K}{K-1} ~ \dfrac{\alpha }{2 \pi}   \label{II} 
\end{align}
substituting \ref{I} into \ref{fReq} gives $\tilde{R} = 1$, then substitute \ref{II} and $\tilde{R}=1$ back into \ref{fQeq} to solve for $\tilde{Q}$, one obtain  
\begin{align}
\tilde{Q} &= 1 + \dfrac{4 \pi}{\alpha \pi K + 2 \alpha}     
\end{align}
substitute the values of $\tilde{Q}, \tilde{R}$ into \ref{freq} and solve for $\tilde{r}$ :
\begin{align}
\tilde{r} &= 1 - \dfrac{2 \pi}{\alpha} \dfrac{K-1}{K} - \dfrac{4 \pi}{\alpha \pi K^{2} + 2 \alpha K} \\
         & = \dfrac{\alpha - 2 \pi}{\alpha} +  \dfrac{2\pi^{2}}{\alpha \pi K +2 \alpha} ~.
\end{align}
Substituting these solutions in Eq.~(\ref{Eq:Eg_QRr}) yields the specialization generalization error Eq.~(\ref{Eq:Eg_spec}). \\

\section*{References}
\bibliography{1st_biblog}

\begin{thebibliography}{45}%
\makeatletter
\providecommand \@ifxundefined [1]{%
 \@ifx{#1\undefined}
}%
\providecommand \@ifnum [1]{%
 \ifnum #1\expandafter \@firstoftwo
 \else \expandafter \@secondoftwo
 \fi
}%
\providecommand \@ifx [1]{%
 \ifx #1\expandafter \@firstoftwo
 \else \expandafter \@secondoftwo
 \fi
}%
\providecommand \natexlab [1]{#1}%
\providecommand \enquote  [1]{``#1''}%
\providecommand \bibnamefont  [1]{#1}%
\providecommand \bibfnamefont [1]{#1}%
\providecommand \citenamefont [1]{#1}%
\providecommand \href@noop [0]{\@secondoftwo}%
\providecommand \href [0]{\begingroup \@sanitize@url \@href}%
\providecommand \@href[1]{\@@startlink{#1}\@@href}%
\providecommand \@@href[1]{\endgroup#1\@@endlink}%
\providecommand \@sanitize@url [0]{\catcode `\\12\catcode `\$12\catcode
  `\&12\catcode `\#12\catcode `\^12\catcode `\_12\catcode `\%12\relax}%
\providecommand \@@startlink[1]{}%
\providecommand \@@endlink[0]{}%
\providecommand \url  [0]{\begingroup\@sanitize@url \@url }%
\providecommand \@url [1]{\endgroup\@href {#1}{\urlprefix }}%
\providecommand \urlprefix  [0]{URL }%
\providecommand \Eprint [0]{\href }%
\providecommand \doibase [0]{https://doi.org/}%
\providecommand \selectlanguage [0]{\@gobble}%
\providecommand \bibinfo  [0]{\@secondoftwo}%
\providecommand \bibfield  [0]{\@secondoftwo}%
\providecommand \translation [1]{[#1]}%
\providecommand \BibitemOpen [0]{}%
\providecommand \bibitemStop [0]{}%
\providecommand \bibitemNoStop [0]{.\EOS\space}%
\providecommand \EOS [0]{\spacefactor3000\relax}%
\providecommand \BibitemShut  [1]{\csname bibitem#1\endcsname}%
\let\auto@bib@innerbib\@empty
\bibitem [{\citenamefont {Engel}\ and\ \citenamefont {Van~den
  Broeck}(2001)}]{engel}%
  \BibitemOpen
  \bibfield  {author} {\bibinfo {author} {\bibfnamefont {A.}~\bibnamefont
  {Engel}}\ and\ \bibinfo {author} {\bibfnamefont {C.}~\bibnamefont {Van~den
  Broeck}},\ }\href {https://doi.org/10.1017/CBO9781139164542} {\emph {\bibinfo
  {title} {Statistical Mechanics of Learning}}}\ (\bibinfo  {publisher}
  {Cambridge University Press},\ \bibinfo {year} {2001})\BibitemShut {NoStop}%
\bibitem [{\citenamefont {Biehl}(2022)}]{book_shallow}%
  \BibitemOpen
  \bibfield  {author} {\bibinfo {author} {\bibfnamefont {M.}~\bibnamefont
  {Biehl}},\ }\href {https://doi.org/10.21827/648c59c1a467e} {\emph {\bibinfo
  {title} {The Shallow and the Deep: A biased introduction to neural networks
  and old school machine learning}}}\ (\bibinfo  {publisher} {University of
  Groningen},\ \bibinfo {year} {2022})\BibitemShut {NoStop}%
\bibitem [{\citenamefont {Carleo}\ \emph {et~al.}(2019)\citenamefont {Carleo},
  \citenamefont {Cirac}, \citenamefont {Cranmer}, \citenamefont {Daudet},
  \citenamefont {Schuld}, \citenamefont {Tishby}, \citenamefont
  {Vogt-Maranto},\ and\ \citenamefont {Zdeborov\'a}}]{phys_Mash_rev}%
  \BibitemOpen
  \bibfield  {author} {\bibinfo {author} {\bibfnamefont {G.}~\bibnamefont
  {Carleo}}, \bibinfo {author} {\bibfnamefont {I.}~\bibnamefont {Cirac}},
  \bibinfo {author} {\bibfnamefont {K.}~\bibnamefont {Cranmer}}, \bibinfo
  {author} {\bibfnamefont {L.}~\bibnamefont {Daudet}}, \bibinfo {author}
  {\bibfnamefont {M.}~\bibnamefont {Schuld}}, \bibinfo {author} {\bibfnamefont
  {N.}~\bibnamefont {Tishby}}, \bibinfo {author} {\bibfnamefont
  {L.}~\bibnamefont {Vogt-Maranto}},\ and\ \bibinfo {author} {\bibfnamefont
  {L.}~\bibnamefont {Zdeborov\'a}},\ }\bibfield  {title} {\bibinfo {title}
  {Machine learning and the physical sciences},\ }\href
  {https://doi.org/10.1103/RevModPhys.91.045002} {\bibfield  {journal}
  {\bibinfo  {journal} {Rev. Mod. Phys.}\ }\textbf {\bibinfo {volume} {91}},\
  \bibinfo {pages} {045002} (\bibinfo {year} {2019})}\BibitemShut {NoStop}%
\bibitem [{\citenamefont {Neal}(1996)}]{Neal1996}%
  \BibitemOpen
  \bibfield  {author} {\bibinfo {author} {\bibfnamefont {R.~M.}\ \bibnamefont
  {Neal}},\ }\href@noop {} {\emph {\bibinfo {title} {Bayesian Learning for
  Neural Networks}}},\ \bibinfo {series} {Lecture Notes in Statistics}, Vol.\
  \bibinfo {volume} {118}\ (\bibinfo  {publisher} {Springer},\ \bibinfo {year}
  {1996})\BibitemShut {NoStop}%
\bibitem [{\citenamefont {Goodfellow}\ \emph {et~al.}(2016)\citenamefont
  {Goodfellow}, \citenamefont {Bengio},\ and\ \citenamefont
  {Courville}}]{goodfellow2016}%
  \BibitemOpen
  \bibfield  {author} {\bibinfo {author} {\bibfnamefont {I.}~\bibnamefont
  {Goodfellow}}, \bibinfo {author} {\bibfnamefont {Y.}~\bibnamefont {Bengio}},\
  and\ \bibinfo {author} {\bibfnamefont {A.}~\bibnamefont {Courville}},\
  }\href@noop {} {\emph {\bibinfo {title} {Deep learning}}}\ (\bibinfo
  {publisher} {MIT press},\ \bibinfo {year} {2016})\BibitemShut {NoStop}%
\bibitem [{\citenamefont {Collins}\ \emph {et~al.}(2021)\citenamefont
  {Collins}, \citenamefont {Dennehy}, \citenamefont {Conboy},\ and\
  \citenamefont {Mikalef}}]{COLLINS2021}%
  \BibitemOpen
  \bibfield  {author} {\bibinfo {author} {\bibfnamefont {C.}~\bibnamefont
  {Collins}}, \bibinfo {author} {\bibfnamefont {D.}~\bibnamefont {Dennehy}},
  \bibinfo {author} {\bibfnamefont {K.}~\bibnamefont {Conboy}},\ and\ \bibinfo
  {author} {\bibfnamefont {P.}~\bibnamefont {Mikalef}},\ }\bibfield  {title}
  {\bibinfo {title} {Artificial intelligence in information systems research: A
  systematic literature review and research agenda},\ }\href
  {https://doi.org/https://doi.org/10.1016/j.ijinfomgt.2021.102383} {\bibfield
  {journal} {\bibinfo  {journal} {International Journal of Information
  Management}\ }\textbf {\bibinfo {volume} {60}},\ \bibinfo {pages} {102383}
  (\bibinfo {year} {2021})}\BibitemShut {NoStop}%
\bibitem [{\citenamefont {Niskanen}\ \emph {et~al.}(2023)\citenamefont
  {Niskanen}, \citenamefont {Sipola},\ and\ \citenamefont
  {Väänänen}}]{niskanen2023}%
  \BibitemOpen
  \bibfield  {author} {\bibinfo {author} {\bibfnamefont {T.}~\bibnamefont
  {Niskanen}}, \bibinfo {author} {\bibfnamefont {T.}~\bibnamefont {Sipola}},\
  and\ \bibinfo {author} {\bibfnamefont {O.}~\bibnamefont {Väänänen}},\
  }\href {https://arxiv.org/abs/2305.04532} {\bibinfo {title} {Latest trends in
  artificial intelligence technology: A scoping review}} (\bibinfo {year}
  {2023}),\ \Eprint {https://arxiv.org/abs/2305.04532} {arXiv:2305.04532
  [cs.LG]} \BibitemShut {NoStop}%
\bibitem [{\citenamefont {Mathew}\ \emph {et~al.}(2021)\citenamefont {Mathew},
  \citenamefont {Amudha},\ and\ \citenamefont {Sivakumari}}]{Deepbook}%
  \BibitemOpen
  \bibfield  {author} {\bibinfo {author} {\bibfnamefont {A.}~\bibnamefont
  {Mathew}}, \bibinfo {author} {\bibfnamefont {P.}~\bibnamefont {Amudha}},\
  and\ \bibinfo {author} {\bibfnamefont {S.}~\bibnamefont {Sivakumari}},\
  }\bibfield  {title} {\bibinfo {title} {Deep learning techniques: An
  overview},\ }in\ \href@noop {} {\emph {\bibinfo {booktitle} {Advanced Machine
  Learning Technologies and Applications}}},\ \bibinfo {editor} {edited by\
  \bibinfo {editor} {\bibfnamefont {A.~E.}\ \bibnamefont {Hassanien}}, \bibinfo
  {editor} {\bibfnamefont {R.}~\bibnamefont {Bhatnagar}},\ and\ \bibinfo
  {editor} {\bibfnamefont {A.}~\bibnamefont {Darwish}}}\ (\bibinfo  {publisher}
  {Springer Singapore},\ \bibinfo {address} {Singapore},\ \bibinfo {year}
  {2021})\ pp.\ \bibinfo {pages} {599--608}\BibitemShut {NoStop}%
\bibitem [{\citenamefont {Dotsenko}(1995)}]{spinNN}%
  \BibitemOpen
  \bibfield  {author} {\bibinfo {author} {\bibfnamefont {V.}~\bibnamefont
  {Dotsenko}},\ }\href {https://doi.org/10.1142/2460} {\emph {\bibinfo {title}
  {An Introduction to the Theory of Spin Glasses and Neural Networks}}}\
  (\bibinfo  {publisher} {WORLD SCIENTIFIC},\ \bibinfo {year} {1995})\ \Eprint
  {https://arxiv.org/abs/https://www.worldscientific.com/doi/pdf/10.1142/2460}
  {https://www.worldscientific.com/doi/pdf/10.1142/2460} \BibitemShut {NoStop}%
\bibitem [{\citenamefont {Bahri}\ \emph {et~al.}(2020)\citenamefont {Bahri},
  \citenamefont {Kadmon}, \citenamefont {Pennington}, \citenamefont
  {Schoenholz}, \citenamefont {Sohl-Dickstein},\ and\ \citenamefont
  {Ganguli}}]{phase_Bahri}%
  \BibitemOpen
  \bibfield  {author} {\bibinfo {author} {\bibfnamefont {Y.}~\bibnamefont
  {Bahri}}, \bibinfo {author} {\bibfnamefont {J.}~\bibnamefont {Kadmon}},
  \bibinfo {author} {\bibfnamefont {J.}~\bibnamefont {Pennington}}, \bibinfo
  {author} {\bibfnamefont {S.~S.}\ \bibnamefont {Schoenholz}}, \bibinfo
  {author} {\bibfnamefont {J.}~\bibnamefont {Sohl-Dickstein}},\ and\ \bibinfo
  {author} {\bibfnamefont {S.}~\bibnamefont {Ganguli}},\ }\bibfield  {title}
  {\bibinfo {title} {Statistical mechanics of deep learning},\ }\href
  {https://doi.org/10.1146/annurev-conmatphys-031119-050745} {\bibfield
  {journal} {\bibinfo  {journal} {Annual Review of Condensed Matter Physics}\
  }\textbf {\bibinfo {volume} {11}},\ \bibinfo {pages} {501} (\bibinfo {year}
  {2020})},\ \Eprint
  {https://arxiv.org/abs/https://doi.org/10.1146/annurev-conmatphys-031119-050745}
  {https://doi.org/10.1146/annurev-conmatphys-031119-050745} \BibitemShut
  {NoStop}%
\bibitem [{\citenamefont {Sompolinsky}\ \emph {et~al.}(1988)\citenamefont
  {Sompolinsky}, \citenamefont {Crisanti},\ and\ \citenamefont
  {Sommers}}]{Sompolinsky1988}%
  \BibitemOpen
  \bibfield  {author} {\bibinfo {author} {\bibfnamefont {H.}~\bibnamefont
  {Sompolinsky}}, \bibinfo {author} {\bibfnamefont {A.}~\bibnamefont
  {Crisanti}},\ and\ \bibinfo {author} {\bibfnamefont {H.~J.}\ \bibnamefont
  {Sommers}},\ }\bibfield  {title} {\bibinfo {title} {Chaos in random neural
  networks},\ }\href {https://doi.org/10.1103/PhysRevLett.61.259} {\bibfield
  {journal} {\bibinfo  {journal} {Physical Review Letters}\ }\textbf {\bibinfo
  {volume} {61}},\ \bibinfo {pages} {259} (\bibinfo {year} {1988})}\BibitemShut
  {NoStop}%
\bibitem [{\citenamefont {Gabri{\'e}}(2020)}]{Gabrie2020}%
  \BibitemOpen
  \bibfield  {author} {\bibinfo {author} {\bibfnamefont {M.}~\bibnamefont
  {Gabri{\'e}}},\ }\bibfield  {title} {\bibinfo {title} {Mean-field inference
  methods for neural networks},\ }\href
  {https://doi.org/10.1088/1751-8121/ab87fd} {\bibfield  {journal} {\bibinfo
  {journal} {Journal of Physics A: Mathematical and Theoretical}\ }\textbf
  {\bibinfo {volume} {53}},\ \bibinfo {pages} {223002} (\bibinfo {year}
  {2020})}\BibitemShut {NoStop}%
\bibitem [{\citenamefont {Mehta}\ \emph {et~al.}(2019)\citenamefont {Mehta},
  \citenamefont {Bukov}, \citenamefont {Wang}, \citenamefont {Day},
  \citenamefont {Richardson}, \citenamefont {Fisher},\ and\ \citenamefont
  {Schwab}}]{MEHTA20191}%
  \BibitemOpen
  \bibfield  {author} {\bibinfo {author} {\bibfnamefont {P.}~\bibnamefont
  {Mehta}}, \bibinfo {author} {\bibfnamefont {M.}~\bibnamefont {Bukov}},
  \bibinfo {author} {\bibfnamefont {C.-H.}\ \bibnamefont {Wang}}, \bibinfo
  {author} {\bibfnamefont {A.~G.}\ \bibnamefont {Day}}, \bibinfo {author}
  {\bibfnamefont {C.}~\bibnamefont {Richardson}}, \bibinfo {author}
  {\bibfnamefont {C.~K.}\ \bibnamefont {Fisher}},\ and\ \bibinfo {author}
  {\bibfnamefont {D.~J.}\ \bibnamefont {Schwab}},\ }\bibfield  {title}
  {\bibinfo {title} {A high-bias, low-variance introduction to machine learning
  for physicists},\ }\href
  {https://doi.org/https://doi.org/10.1016/j.physrep.2019.03.001} {\bibfield
  {journal} {\bibinfo  {journal} {Physics Reports}\ }\textbf {\bibinfo {volume}
  {810}},\ \bibinfo {pages} {1} (\bibinfo {year} {2019})},\ \bibinfo {note} {a
  high-bias, low-variance introduction to Machine Learning for
  physicists}\BibitemShut {NoStop}%
\bibitem [{\citenamefont {Watkin}\ \emph {et~al.}(1993)\citenamefont {Watkin},
  \citenamefont {Rau},\ and\ \citenamefont {Biehl}}]{RevModPhys.65.499}%
  \BibitemOpen
  \bibfield  {author} {\bibinfo {author} {\bibfnamefont {T.~L.~H.}\
  \bibnamefont {Watkin}}, \bibinfo {author} {\bibfnamefont {A.}~\bibnamefont
  {Rau}},\ and\ \bibinfo {author} {\bibfnamefont {M.}~\bibnamefont {Biehl}},\
  }\bibfield  {title} {\bibinfo {title} {The statistical mechanics of learning
  a rule},\ }\href {https://doi.org/10.1103/RevModPhys.65.499} {\bibfield
  {journal} {\bibinfo  {journal} {Rev. Mod. Phys.}\ }\textbf {\bibinfo {volume}
  {65}},\ \bibinfo {pages} {499} (\bibinfo {year} {1993})}\BibitemShut
  {NoStop}%
\bibitem [{\citenamefont {{Gardner, E.}}\ \emph {et~al.}(1987)\citenamefont
  {{Gardner, E.}}, \citenamefont {{Derrida, B.}},\ and\ \citenamefont
  {{Mottishaw, P.}}}]{gardner}%
  \BibitemOpen
  \bibfield  {author} {\bibinfo {author} {\bibnamefont {{Gardner, E.}}},
  \bibinfo {author} {\bibnamefont {{Derrida, B.}}},\ and\ \bibinfo {author}
  {\bibnamefont {{Mottishaw, P.}}},\ }\bibfield  {title} {\bibinfo {title}
  {Zero temperature parallel dynamics for infinite range spin glasses and
  neural networks},\ }\href {https://doi.org/10.1051/jphys:01987004805074100}
  {\bibfield  {journal} {\bibinfo  {journal} {J. Phys. France}\ }\textbf
  {\bibinfo {volume} {48}},\ \bibinfo {pages} {741} (\bibinfo {year}
  {1987})}\BibitemShut {NoStop}%
\bibitem [{\citenamefont {Seung}\ \emph {et~al.}(1992)\citenamefont {Seung},
  \citenamefont {Sompolinsky},\ and\ \citenamefont {Tishby}}]{Tishby}%
  \BibitemOpen
  \bibfield  {author} {\bibinfo {author} {\bibfnamefont {H.~S.}\ \bibnamefont
  {Seung}}, \bibinfo {author} {\bibfnamefont {H.}~\bibnamefont {Sompolinsky}},\
  and\ \bibinfo {author} {\bibfnamefont {N.}~\bibnamefont {Tishby}},\
  }\bibfield  {title} {\bibinfo {title} {Statistical mechanics of learning from
  examples},\ }\href {https://doi.org/10.1103/PhysRevA.45.6056} {\bibfield
  {journal} {\bibinfo  {journal} {Phys. Rev. A}\ }\textbf {\bibinfo {volume}
  {45}},\ \bibinfo {pages} {6056} (\bibinfo {year} {1992})}\BibitemShut
  {NoStop}%
\bibitem [{\citenamefont {Gardner}(1988)}]{Gardner1988}%
  \BibitemOpen
  \bibfield  {author} {\bibinfo {author} {\bibfnamefont {E.}~\bibnamefont
  {Gardner}},\ }\bibfield  {title} {\bibinfo {title} {The space of interactions
  in neural network models},\ }\href
  {https://doi.org/10.1088/0305-4470/21/1/030} {\bibfield  {journal} {\bibinfo
  {journal} {Journal of Physics A: Mathematical and General}\ }\textbf
  {\bibinfo {volume} {21}},\ \bibinfo {pages} {257} (\bibinfo {year}
  {1988})}\BibitemShut {NoStop}%
\bibitem [{\citenamefont {Levin}\ \emph {et~al.}(1990)\citenamefont {Levin},
  \citenamefont {Tishby},\ and\ \citenamefont {Solla}}]{Levin}%
  \BibitemOpen
  \bibfield  {author} {\bibinfo {author} {\bibfnamefont {E.}~\bibnamefont
  {Levin}}, \bibinfo {author} {\bibfnamefont {N.}~\bibnamefont {Tishby}},\ and\
  \bibinfo {author} {\bibfnamefont {S.}~\bibnamefont {Solla}},\ }\bibfield
  {title} {\bibinfo {title} {A statistical approach to learning and
  generalization in layered neural networks},\ }\href
  {https://doi.org/10.1109/5.58339} {\bibfield  {journal} {\bibinfo  {journal}
  {Proceedings of the IEEE}\ }\textbf {\bibinfo {volume} {78}},\ \bibinfo
  {pages} {1568} (\bibinfo {year} {1990})}\BibitemShut {NoStop}%
\bibitem [{\citenamefont {Rosen-Zvi}\ \emph {et~al.}(2001)\citenamefont
  {Rosen-Zvi}, \citenamefont {Engel},\ and\ \citenamefont
  {Kanter}}]{PhysRevLett.87.078101}%
  \BibitemOpen
  \bibfield  {author} {\bibinfo {author} {\bibfnamefont {M.}~\bibnamefont
  {Rosen-Zvi}}, \bibinfo {author} {\bibfnamefont {A.}~\bibnamefont {Engel}},\
  and\ \bibinfo {author} {\bibfnamefont {I.}~\bibnamefont {Kanter}},\
  }\bibfield  {title} {\bibinfo {title} {Multilayer neural networks with
  extensively many hidden units},\ }\href
  {https://doi.org/10.1103/PhysRevLett.87.078101} {\bibfield  {journal}
  {\bibinfo  {journal} {Phys. Rev. Lett.}\ }\textbf {\bibinfo {volume} {87}},\
  \bibinfo {pages} {078101} (\bibinfo {year} {2001})}\BibitemShut {NoStop}%
\bibitem [{\citenamefont {Urbanczik}(1995)}]{RUrbanczik_1995}%
  \BibitemOpen
  \bibfield  {author} {\bibinfo {author} {\bibfnamefont {R.}~\bibnamefont
  {Urbanczik}},\ }\bibfield  {title} {\bibinfo {title} {A fully connected
  committee machine learning unrealizable rules},\ }\href
  {https://doi.org/10.1088/0305-4470/28/24/010} {\bibfield  {journal} {\bibinfo
   {journal} {Journal of Physics A: Mathematical and General}\ }\textbf
  {\bibinfo {volume} {28}},\ \bibinfo {pages} {7097} (\bibinfo {year}
  {1995})}\BibitemShut {NoStop}%
\bibitem [{\citenamefont {Saad}\ and\ \citenamefont
  {Solla}(1995{\natexlab{a}})}]{saad}%
  \BibitemOpen
  \bibfield  {author} {\bibinfo {author} {\bibfnamefont {D.}~\bibnamefont
  {Saad}}\ and\ \bibinfo {author} {\bibfnamefont {S.~A.}\ \bibnamefont
  {Solla}},\ }\bibfield  {title} {\bibinfo {title} {On-line learning in soft
  committee machines},\ }\href {https://doi.org/10.1103/PhysRevE.52.4225}
  {\bibfield  {journal} {\bibinfo  {journal} {Phys. Rev. E}\ }\textbf {\bibinfo
  {volume} {52}},\ \bibinfo {pages} {4225} (\bibinfo {year}
  {1995}{\natexlab{a}})}\BibitemShut {NoStop}%
\bibitem [{\citenamefont {Saad}\ and\ \citenamefont
  {Solla}(1995{\natexlab{b}})}]{Saad95prl}%
  \BibitemOpen
  \bibfield  {author} {\bibinfo {author} {\bibfnamefont {D.}~\bibnamefont
  {Saad}}\ and\ \bibinfo {author} {\bibfnamefont {S.~A.}\ \bibnamefont
  {Solla}},\ }\bibfield  {title} {\bibinfo {title} {Exact solution for on-line
  learning in multilayer neural networks},\ }\href
  {https://doi.org/10.1103/PhysRevLett.74.4337} {\bibfield  {journal} {\bibinfo
   {journal} {Phys. Rev. Lett.}\ }\textbf {\bibinfo {volume} {74}},\ \bibinfo
  {pages} {4337} (\bibinfo {year} {1995}{\natexlab{b}})}\BibitemShut {NoStop}%
\bibitem [{\citenamefont {Nair}\ and\ \citenamefont {Hinton}(2010)}]{nair2010}%
  \BibitemOpen
  \bibfield  {author} {\bibinfo {author} {\bibfnamefont {V.}~\bibnamefont
  {Nair}}\ and\ \bibinfo {author} {\bibfnamefont {G.~E.}\ \bibnamefont
  {Hinton}},\ }\bibfield  {title} {\bibinfo {title} {Rectified linear units
  improve restricted boltzmann machines},\ }in\ \href@noop {} {\emph {\bibinfo
  {booktitle} {ICML 2010}}}\ (\bibinfo {year} {2010})\ pp.\ \bibinfo {pages}
  {807--814}\BibitemShut {NoStop}%
\bibitem [{\citenamefont {Belkin}\ \emph {et~al.}(2019)\citenamefont {Belkin},
  \citenamefont {Hsu}, \citenamefont {Ma},\ and\ \citenamefont
  {Mandal}}]{doubledecent}%
  \BibitemOpen
  \bibfield  {author} {\bibinfo {author} {\bibfnamefont {M.}~\bibnamefont
  {Belkin}}, \bibinfo {author} {\bibfnamefont {D.}~\bibnamefont {Hsu}},
  \bibinfo {author} {\bibfnamefont {S.}~\bibnamefont {Ma}},\ and\ \bibinfo
  {author} {\bibfnamefont {S.}~\bibnamefont {Mandal}},\ }\bibfield  {title}
  {\bibinfo {title} {Reconciling modern machine learning practice and the
  classical bias variance trade off},\ }\href
  {https://doi.org/10.1073/pnas.1903070116} {\bibfield  {journal} {\bibinfo
  {journal} {Proceedings of the National Academy of Sciences}\ }\textbf
  {\bibinfo {volume} {116}},\ \bibinfo {pages} {15849} (\bibinfo {year}
  {2019})},\ \Eprint
  {https://arxiv.org/abs/https://www.pnas.org/doi/pdf/10.1073/pnas.1903070116}
  {https://www.pnas.org/doi/pdf/10.1073/pnas.1903070116} \BibitemShut {NoStop}%
\bibitem [{\citenamefont {Rosen-Zvi}\ \emph {et~al.}(2002)\citenamefont
  {Rosen-Zvi}, \citenamefont {Engel},\ and\ \citenamefont {Kanter}}]{Rosen}%
  \BibitemOpen
  \bibfield  {author} {\bibinfo {author} {\bibfnamefont {M.}~\bibnamefont
  {Rosen-Zvi}}, \bibinfo {author} {\bibfnamefont {A.}~\bibnamefont {Engel}},\
  and\ \bibinfo {author} {\bibfnamefont {I.}~\bibnamefont {Kanter}},\
  }\bibfield  {title} {\bibinfo {title} {Generalization and capacity of
  extensively large two-layered perceptrons},\ }\href
  {https://doi.org/10.1103/PhysRevE.66.036138} {\bibfield  {journal} {\bibinfo
  {journal} {Phys. Rev. E}\ }\textbf {\bibinfo {volume} {66}},\ \bibinfo
  {pages} {036138} (\bibinfo {year} {2002})}\BibitemShut {NoStop}%
\bibitem [{\citenamefont {Lee}\ \emph {et~al.}(2018)\citenamefont {Lee},
  \citenamefont {Bahri}, \citenamefont {Novak}, \citenamefont {Schoenholz},
  \citenamefont {Pennington},\ and\ \citenamefont {Sohl-Dickstein}}]{lee2018}%
  \BibitemOpen
  \bibfield  {author} {\bibinfo {author} {\bibfnamefont {J.}~\bibnamefont
  {Lee}}, \bibinfo {author} {\bibfnamefont {Y.}~\bibnamefont {Bahri}}, \bibinfo
  {author} {\bibfnamefont {R.}~\bibnamefont {Novak}}, \bibinfo {author}
  {\bibfnamefont {S.~S.}\ \bibnamefont {Schoenholz}}, \bibinfo {author}
  {\bibfnamefont {J.}~\bibnamefont {Pennington}},\ and\ \bibinfo {author}
  {\bibfnamefont {J.}~\bibnamefont {Sohl-Dickstein}},\ }\href
  {https://arxiv.org/abs/1711.00165} {\bibinfo {title} {Deep neural networks as
  gaussian processes}} (\bibinfo {year} {2018}),\ \Eprint
  {https://arxiv.org/abs/1711.00165} {arXiv:1711.00165 [stat.ML]} \BibitemShut
  {NoStop}%
\bibitem [{\citenamefont {Jacot}\ \emph {et~al.}(2018)\citenamefont {Jacot},
  \citenamefont {Gabriel},\ and\ \citenamefont {Hongler}}]{Jacot2018}%
  \BibitemOpen
  \bibfield  {author} {\bibinfo {author} {\bibfnamefont {A.}~\bibnamefont
  {Jacot}}, \bibinfo {author} {\bibfnamefont {F.}~\bibnamefont {Gabriel}},\
  and\ \bibinfo {author} {\bibfnamefont {C.}~\bibnamefont {Hongler}},\
  }\bibfield  {title} {\bibinfo {title} {Neural tangent kernel: Convergence and
  generalization in neural networks},\ }in\ \href@noop {} {\emph {\bibinfo
  {booktitle} {Advances in Neural Information Processing Systems (NeurIPS)}}},\
  Vol.~\bibinfo {volume} {31}\ (\bibinfo {year} {2018})\ pp.\ \bibinfo {pages}
  {8571--8580}\BibitemShut {NoStop}%
\bibitem [{\citenamefont {Arora}\ \emph {et~al.}(2019)\citenamefont {Arora},
  \citenamefont {Du}, \citenamefont {Hu}, \citenamefont {Li}, \citenamefont
  {Salakhutdinov},\ and\ \citenamefont {Wang}}]{Arora2019}%
  \BibitemOpen
  \bibfield  {author} {\bibinfo {author} {\bibfnamefont {S.}~\bibnamefont
  {Arora}}, \bibinfo {author} {\bibfnamefont {S.~S.}\ \bibnamefont {Du}},
  \bibinfo {author} {\bibfnamefont {W.}~\bibnamefont {Hu}}, \bibinfo {author}
  {\bibfnamefont {Z.}~\bibnamefont {Li}}, \bibinfo {author} {\bibfnamefont
  {R.}~\bibnamefont {Salakhutdinov}},\ and\ \bibinfo {author} {\bibfnamefont
  {R.}~\bibnamefont {Wang}},\ }\bibfield  {title} {\bibinfo {title} {On exact
  computation with an infinitely wide neural net},\ }in\ \href@noop {} {\emph
  {\bibinfo {booktitle} {Advances in Neural Information Processing Systems
  (NeurIPS)}}},\ Vol.~\bibinfo {volume} {32}\ (\bibinfo {year}
  {2019})\BibitemShut {NoStop}%
\bibitem [{\citenamefont {Allen-Zhu}\ \emph {et~al.}(2019)\citenamefont
  {Allen-Zhu}, \citenamefont {Li},\ and\ \citenamefont {Song}}]{AllenZhu2019}%
  \BibitemOpen
  \bibfield  {author} {\bibinfo {author} {\bibfnamefont {Z.}~\bibnamefont
  {Allen-Zhu}}, \bibinfo {author} {\bibfnamefont {Y.}~\bibnamefont {Li}},\ and\
  \bibinfo {author} {\bibfnamefont {Z.}~\bibnamefont {Song}},\ }\bibfield
  {title} {\bibinfo {title} {A convergence theory for deep learning via
  over-parameterization},\ }in\ \href@noop {} {\emph {\bibinfo {booktitle}
  {Proceedings of the 36th International Conference on Machine Learning
  (ICML)}}}\ (\bibinfo {year} {2019})\ pp.\ \bibinfo {pages}
  {242--252}\BibitemShut {NoStop}%
\bibitem [{\citenamefont {Advani}\ \emph {et~al.}(2020)\citenamefont {Advani},
  \citenamefont {Saxe},\ and\ \citenamefont {Sompolinsky}}]{Advani2020}%
  \BibitemOpen
  \bibfield  {author} {\bibinfo {author} {\bibfnamefont {M.~S.}\ \bibnamefont
  {Advani}}, \bibinfo {author} {\bibfnamefont {A.~M.}\ \bibnamefont {Saxe}},\
  and\ \bibinfo {author} {\bibfnamefont {H.}~\bibnamefont {Sompolinsky}},\
  }\bibfield  {title} {\bibinfo {title} {High-dimensional dynamics of
  generalization error in neural networks},\ }\href
  {https://doi.org/10.1016/j.neunet.2020.09.010} {\bibfield  {journal}
  {\bibinfo  {journal} {Neural Networks}\ }\textbf {\bibinfo {volume} {132}},\
  \bibinfo {pages} {428} (\bibinfo {year} {2020})}\BibitemShut {NoStop}%
\bibitem [{\citenamefont {Li}\ and\ \citenamefont
  {Sompolinsky}(2021)}]{Li2021}%
  \BibitemOpen
  \bibfield  {author} {\bibinfo {author} {\bibfnamefont {Q.}~\bibnamefont
  {Li}}\ and\ \bibinfo {author} {\bibfnamefont {H.}~\bibnamefont
  {Sompolinsky}},\ }\bibfield  {title} {\bibinfo {title} {Statistical mechanics
  of deep linear neural networks: The backpropagating kernel renormalization},\
  }\href {https://doi.org/10.1103/PhysRevX.11.031059} {\bibfield  {journal}
  {\bibinfo  {journal} {Physical Review X}\ }\textbf {\bibinfo {volume} {11}},\
  \bibinfo {pages} {031059} (\bibinfo {year} {2021})}\BibitemShut {NoStop}%
\bibitem [{\citenamefont {Schwarze}\ and\ \citenamefont
  {Hertz}(1993{\natexlab{a}})}]{H.Schwarze_1993}%
  \BibitemOpen
  \bibfield  {author} {\bibinfo {author} {\bibfnamefont {H.}~\bibnamefont
  {Schwarze}}\ and\ \bibinfo {author} {\bibfnamefont {J.}~\bibnamefont
  {Hertz}},\ }\bibfield  {title} {\bibinfo {title} {Generalization in fully
  connected committee machines},\ }\href
  {https://doi.org/10.1209/0295-5075/21/7/012} {\bibfield  {journal} {\bibinfo
  {journal} {Europhysics Letters}\ }\textbf {\bibinfo {volume} {21}},\ \bibinfo
  {pages} {785} (\bibinfo {year} {1993}{\natexlab{a}})}\BibitemShut {NoStop}%
\bibitem [{\citenamefont {Schwarze}\ and\ \citenamefont
  {Hertz}(1993{\natexlab{b}})}]{HSchwarze2_1993}%
  \BibitemOpen
  \bibfield  {author} {\bibinfo {author} {\bibfnamefont {H.}~\bibnamefont
  {Schwarze}}\ and\ \bibinfo {author} {\bibfnamefont {J.}~\bibnamefont
  {Hertz}},\ }\bibfield  {title} {\bibinfo {title} {Learning from examples in
  fully connected committee machines},\ }\href
  {https://doi.org/10.1088/0305-4470/26/19/024} {\bibfield  {journal} {\bibinfo
   {journal} {Journal of Physics A: Mathematical and General}\ }\textbf
  {\bibinfo {volume} {26}},\ \bibinfo {pages} {4919} (\bibinfo {year}
  {1993}{\natexlab{b}})}\BibitemShut {NoStop}%
\bibitem [{\citenamefont {Schwarze}(1993)}]{HSchwarze_1993}%
  \BibitemOpen
  \bibfield  {author} {\bibinfo {author} {\bibfnamefont {H.}~\bibnamefont
  {Schwarze}},\ }\bibfield  {title} {\bibinfo {title} {Learning a rule in a
  multilayer neural network},\ }\href
  {https://doi.org/10.1088/0305-4470/26/21/017} {\bibfield  {journal} {\bibinfo
   {journal} {Journal of Physics A: Mathematical and General}\ }\textbf
  {\bibinfo {volume} {26}},\ \bibinfo {pages} {5781} (\bibinfo {year}
  {1993})}\BibitemShut {NoStop}%
\bibitem [{\citenamefont {Ahr}\ \emph {et~al.}(1999)\citenamefont {Ahr},
  \citenamefont {Biehl},\ and\ \citenamefont {Urbanczik}}]{Ahr_1999}%
  \BibitemOpen
  \bibfield  {author} {\bibinfo {author} {\bibfnamefont {M.}~\bibnamefont
  {Ahr}}, \bibinfo {author} {\bibfnamefont {M.}~\bibnamefont {Biehl}},\ and\
  \bibinfo {author} {\bibfnamefont {R.}~\bibnamefont {Urbanczik}},\ }\bibfield
  {title} {\bibinfo {title} {Statistical physics and practical training of
  soft-committee machines},\ }\href {https://doi.org/10.1007/s100510050889}
  {\bibfield  {journal} {\bibinfo  {journal} {The European Physical Journal B}\
  }\textbf {\bibinfo {volume} {10}},\ \bibinfo {pages} {583} (\bibinfo {year}
  {1999})}\BibitemShut {NoStop}%
\bibitem [{\citenamefont {M{\'e}zard}\ \emph {et~al.}(1987)\citenamefont
  {M{\'e}zard}, \citenamefont {Parisi},\ and\ \citenamefont
  {Virasoro}}]{Mezard1987}%
  \BibitemOpen
  \bibfield  {author} {\bibinfo {author} {\bibfnamefont {M.}~\bibnamefont
  {M{\'e}zard}}, \bibinfo {author} {\bibfnamefont {G.}~\bibnamefont {Parisi}},\
  and\ \bibinfo {author} {\bibfnamefont {M.}~\bibnamefont {Virasoro}},\
  }\href@noop {} {\emph {\bibinfo {title} {Spin Glass Theory and Beyond}}}\
  (\bibinfo  {publisher} {World Scientific},\ \bibinfo {year}
  {1987})\BibitemShut {NoStop}%
\bibitem [{\citenamefont {Oostwal}\ \emph {et~al.}(2021)\citenamefont
  {Oostwal}, \citenamefont {Straat},\ and\ \citenamefont {Biehl}}]{Oostwal}%
  \BibitemOpen
  \bibfield  {author} {\bibinfo {author} {\bibfnamefont {E.}~\bibnamefont
  {Oostwal}}, \bibinfo {author} {\bibfnamefont {M.}~\bibnamefont {Straat}},\
  and\ \bibinfo {author} {\bibfnamefont {M.}~\bibnamefont {Biehl}},\ }\bibfield
   {title} {\bibinfo {title} {Hidden unit specialization in layered neural
  networks: Relu vs. sigmoidal activation},\ }\href
  {https://doi.org/10.1016/j.physa.2020.125517} {\bibfield  {journal} {\bibinfo
   {journal} {Physica A: Statistical Mechanics and its Applications}\ }\textbf
  {\bibinfo {volume} {564}},\ \bibinfo {pages} {125517} (\bibinfo {year}
  {2021})}\BibitemShut {NoStop}%
\bibitem [{\citenamefont {Biehl}\ \emph {et~al.}(1998)\citenamefont {Biehl},
  \citenamefont {Schlosser},\ and\ \citenamefont {Ahr}}]{Biehl_1998}%
  \BibitemOpen
  \bibfield  {author} {\bibinfo {author} {\bibfnamefont {M.}~\bibnamefont
  {Biehl}}, \bibinfo {author} {\bibfnamefont {E.}~\bibnamefont {Schlosser}},\
  and\ \bibinfo {author} {\bibfnamefont {M.}~\bibnamefont {Ahr}},\ }\bibfield
  {title} {\bibinfo {title} {Phase transitions in soft-committee machines},\
  }\href {https://doi.org/10.1209/epl/i1998-00466-6} {\bibfield  {journal}
  {\bibinfo  {journal} {Europhysics Letters}\ }\textbf {\bibinfo {volume}
  {44}},\ \bibinfo {pages} {261} (\bibinfo {year} {1998})}\BibitemShut
  {NoStop}%
\bibitem [{\citenamefont {Richert}\ \emph {et~al.}(2022)\citenamefont
  {Richert}, \citenamefont {Worschech},\ and\ \citenamefont
  {Rosenow}}]{Richert_2022}%
  \BibitemOpen
  \bibfield  {author} {\bibinfo {author} {\bibfnamefont {F.}~\bibnamefont
  {Richert}}, \bibinfo {author} {\bibfnamefont {R.}~\bibnamefont {Worschech}},\
  and\ \bibinfo {author} {\bibfnamefont {B.}~\bibnamefont {Rosenow}},\
  }\bibfield  {title} {\bibinfo {title} {Soft mode in the dynamics of
  over-realizable online learning for soft committee machines},\ }\bibfield
  {journal} {\bibinfo  {journal} {Physical Review E}\ }\textbf {\bibinfo
  {volume} {105}},\ \href {https://doi.org/10.1103/physreve.105.l052302}
  {10.1103/physreve.105.l052302} (\bibinfo {year} {2022})\BibitemShut {NoStop}%
\bibitem [{\citenamefont {Zeiler}\ \emph {et~al.}(2013)\citenamefont {Zeiler},
  \citenamefont {Ranzato}, \citenamefont {Monga}, \citenamefont {Mao},
  \citenamefont {Yang}, \citenamefont {Le}, \citenamefont {Nguyen},
  \citenamefont {Senior}, \citenamefont {Vanhoucke}, \citenamefont {Dean},\
  and\ \citenamefont {Hinton}}]{relu_speech}%
  \BibitemOpen
  \bibfield  {author} {\bibinfo {author} {\bibfnamefont {M.}~\bibnamefont
  {Zeiler}}, \bibinfo {author} {\bibfnamefont {M.}~\bibnamefont {Ranzato}},
  \bibinfo {author} {\bibfnamefont {R.}~\bibnamefont {Monga}}, \bibinfo
  {author} {\bibfnamefont {M.}~\bibnamefont {Mao}}, \bibinfo {author}
  {\bibfnamefont {K.}~\bibnamefont {Yang}}, \bibinfo {author} {\bibfnamefont
  {Q.}~\bibnamefont {Le}}, \bibinfo {author} {\bibfnamefont {P.}~\bibnamefont
  {Nguyen}}, \bibinfo {author} {\bibfnamefont {A.}~\bibnamefont {Senior}},
  \bibinfo {author} {\bibfnamefont {V.}~\bibnamefont {Vanhoucke}}, \bibinfo
  {author} {\bibfnamefont {J.}~\bibnamefont {Dean}},\ and\ \bibinfo {author}
  {\bibfnamefont {G.}~\bibnamefont {Hinton}},\ }\bibfield  {title} {\bibinfo
  {title} {On rectified linear units for speech processing},\ }in\ \href@noop
  {} {\emph {\bibinfo {booktitle} {38th International Conference on Acoustics,
  Speech and Signal Processing (ICASSP)}}}\ (\bibinfo {address} {Vancouver},\
  \bibinfo {year} {2013})\BibitemShut {NoStop}%
\bibitem [{\citenamefont {Xu}\ \emph {et~al.}(2015)\citenamefont {Xu},
  \citenamefont {Wang}, \citenamefont {Chen},\ and\ \citenamefont
  {Li}}]{relu_empirical}%
  \BibitemOpen
  \bibfield  {author} {\bibinfo {author} {\bibfnamefont {B.}~\bibnamefont
  {Xu}}, \bibinfo {author} {\bibfnamefont {N.}~\bibnamefont {Wang}}, \bibinfo
  {author} {\bibfnamefont {T.}~\bibnamefont {Chen}},\ and\ \bibinfo {author}
  {\bibfnamefont {M.}~\bibnamefont {Li}},\ }\href@noop {} {\bibinfo {title}
  {Empirical evaluation of rectified activations in convolutional network}}
  (\bibinfo {year} {2015}),\ \Eprint {https://arxiv.org/abs/1505.00853}
  {arXiv:1505.00853 [cs.LG]} \BibitemShut {NoStop}%
\bibitem [{\citenamefont {B\"os}\ \emph {et~al.}(1993)\citenamefont {B\"os},
  \citenamefont {Kinzel},\ and\ \citenamefont {Opper}}]{PhysRevE.47.1384}%
  \BibitemOpen
  \bibfield  {author} {\bibinfo {author} {\bibfnamefont {S.}~\bibnamefont
  {B\"os}}, \bibinfo {author} {\bibfnamefont {W.}~\bibnamefont {Kinzel}},\ and\
  \bibinfo {author} {\bibfnamefont {M.}~\bibnamefont {Opper}},\ }\bibfield
  {title} {\bibinfo {title} {Generalization ability of perceptrons with
  continuous outputs},\ }\href {https://doi.org/10.1103/PhysRevE.47.1384}
  {\bibfield  {journal} {\bibinfo  {journal} {Phys. Rev. E}\ }\textbf {\bibinfo
  {volume} {47}},\ \bibinfo {pages} {1384} (\bibinfo {year}
  {1993})}\BibitemShut {NoStop}%
\bibitem [{\citenamefont {Andrecut}(2018)}]{Andrecut_2018}%
  \BibitemOpen
  \bibfield  {author} {\bibinfo {author} {\bibfnamefont {M.}~\bibnamefont
  {Andrecut}},\ }\bibfield  {title} {\bibinfo {title} {High-dimensional vector
  semantics},\ }\href {https://doi.org/10.1142/s0129183118500158} {\bibfield
  {journal} {\bibinfo  {journal} {International Journal of Modern Physics C}\
  }\textbf {\bibinfo {volume} {29}},\ \bibinfo {pages} {1850015} (\bibinfo
  {year} {2018})}\BibitemShut {NoStop}%
\bibitem [{\citenamefont {Barbier}\ \emph {et~al.}(2025)\citenamefont
  {Barbier}, \citenamefont {Camilli}, \citenamefont {Nguyen}, \citenamefont
  {Pastore},\ and\ \citenamefont {Skerk}}]{Barbier25}%
  \BibitemOpen
  \bibfield  {author} {\bibinfo {author} {\bibfnamefont {J.}~\bibnamefont
  {Barbier}}, \bibinfo {author} {\bibfnamefont {F.}~\bibnamefont {Camilli}},
  \bibinfo {author} {\bibfnamefont {M.-T.}\ \bibnamefont {Nguyen}}, \bibinfo
  {author} {\bibfnamefont {M.}~\bibnamefont {Pastore}},\ and\ \bibinfo {author}
  {\bibfnamefont {R.}~\bibnamefont {Skerk}},\ }\href
  {https://doi.org/10.48550/ARXIV.2510.24616} {\bibinfo {title} {Statistical
  physics of deep learning: Optimal learning of a multi-layer perceptron near
  interpolation}} (\bibinfo {year} {2025})\BibitemShut {NoStop}%
\bibitem [{\citenamefont {Citton}\ \emph {et~al.}(2025)\citenamefont {Citton},
  \citenamefont {Richert},\ and\ \citenamefont {Biehl}}]{Otavio2025}%
  \BibitemOpen
  \bibfield  {author} {\bibinfo {author} {\bibfnamefont {O.}~\bibnamefont
  {Citton}}, \bibinfo {author} {\bibfnamefont {F.}~\bibnamefont {Richert}},\
  and\ \bibinfo {author} {\bibfnamefont {M.}~\bibnamefont {Biehl}},\ }\bibfield
   {title} {\bibinfo {title} {Phase transition analysis for shallow neural
  networks with arbitrary activation functions},\ }\href
  {https://doi.org/https://doi.org/10.1016/j.physa.2025.130356} {\bibfield
  {journal} {\bibinfo  {journal} {Physica A: Statistical Mechanics and its
  Applications}\ }\textbf {\bibinfo {volume} {660}},\ \bibinfo {pages} {130356}
  (\bibinfo {year} {2025})}\BibitemShut {NoStop}%
\end{thebibliography}%
\nocite{*}

\end{document}